\documentclass[3p,times,12pt]{elsarticle}

\usepackage{amssymb}
\usepackage[figuresright]{rotating}
\usepackage{color} %%%YS
\usepackage{url}
\newcommand{\be}{\begin{equation}}
\newcommand{\ee}{\end{equation}}
\begin{document}

\begin{frontmatter}



%%%-YS \dochead{}
%% e.g. \dochead{17th International Conference on Dynamical Processes in Excited States of Solids}

\title{Decay of Benjamin--Ono solitons under the influence of dissipation} %


\author{Roger H. Grimshaw$^{a)}$, Noel F. Smyth$^{b)}$, Yury A. Stepanyants$^{c, d)}$
\footnote{Corresponding author,
e-mail: Yury.Stepanyants@usq.edu.au}} %

\address{\vspace*{0.5cm} $^{a)}$ University College London, UK; \\%
$^{b)}$ School of Mathematics, University of Edinburgh,\\
James Clerk Maxwell Building, The King's Buildings,\\
Peter Guthrie Tait Road, Edinburgh, Scotland, U.K., EH9 3FD; \\%
$^{c)}$ Faculty of Health, Engineering and Sciences,\\ University
of Southern Queensland, Toowoomba, QLD, 4350, Australia and \\
$^{d)}$ Department of Applied Mathematics, Nizhny Novgorod State
Technical University n.a. R.E. Alekseev, Nizhny Novgorod, Russia.
\\}

\begin{abstract}
The adiabatic decay of Benjamin--Ono algebraic solitons is studied
when the influence of various types of small dissipation and
radiative losses due to large scale Coriolis dispersion are taken
into consideration. The physically most important dissipations are
studied, Rayleigh and Reynolds dissipation, Landau damping,
dissipation in a laminar boundary layer and Chezy friction on a
rough bottom. The decay laws for the soliton parameters, that is
amplitude, velocity and width, are found in analytical form and
are compared with the results of direct numerical modelling.
\end{abstract}

\begin{keyword}
internal wave \sep Benjamin--Ono equation \sep soliton \sep
Rayleigh dissipation \sep Reynolds dissipation \sep Landau damping
\sep Chezy friction \sep rotating fluid \sep adiabatic decay

%\PACS 03.40.Gc \sep 03.40.Kf \sep 04.20.Jb \sep 63.20.Ry


\end{keyword}

\end{frontmatter}

% \linenumbers

\newpage%
\section{Introduction}
\label{Sect1}%

The propagation of small amplitude long waves in a stratified
fluid consisting of a relatively thin layer overlying a very deep
passive layer is described by the well-known Benjamin--Ono (BO)
equation \cite{Ablowitz-1981,Apel-2007,Grimshaw-2001}
\begin{equation}%
\label{Eq01}%
\frac{\partial u}{\partial t} + \alpha u \frac{\partial
u}{\partial x} + \frac{\beta}{\pi} \frac{\partial^2}{\partial
x^2}\,\wp\!\!\int\limits_{-\infty}^{+\infty}\frac{u(\xi, t)}{\xi -
x}d\xi = 0.
\end{equation}
Here, $u(x, t)$ is the perturbation of a pycnocline (a layer with
a constant density) and $\alpha$ and $\beta$ are parameters which
depend on the particular stratification (for details see
\cite{Apel-2007,Grimshaw-2001}).  The symbol $\wp$ denotes the
principal value of the integral.  The BO equation (\ref{Eq01}) is
set in a coordinate frame moving with the speed $c$ of linear
long waves.  In particular, the BO equation has the algebraic soliton
solution \cite{Ablowitz-1981}
\begin{equation}%
\label{Eq02}%
u(x, t) = \frac{A}{1 + (x - Vt)^2/\Delta^2}.
\end{equation}
Here, $A$ is the soliton amplitude, $V = \alpha A/4$ is its
velocity in the Galilean coordinate frame moving with the speed $c$
with respect to an immovable observer and $\Delta = 4\beta/\alpha
A$ is its characteristic width.

The BO equation is similar to the classic Korteweg-de Vries (KdV)
equation for weakly nonlinear long waves in a shallow fluid
\begin{equation}
\label{e:kdv} %
\frac{\partial u}{\partial t} + \alpha
u\frac{\partial u}{\partial x} +
\beta\frac{\partial^{3}u}{\partial x^{3}} = 0.
\end{equation}
In particular, both are completely integrable \cite{Ablowitz-1981}
and have families of periodic wave solutions, one limit of which
is the solitary wave solution. A specific feature of these
solitary wave solutions is that they are robust and restore their
parameters (amplitude, shape etc.) after collisions with each
other, so that they are termed solitons \cite{Zabusky-1965}, the
only relic of the interaction being a phase shift. They are also
stable under arbitrary localised perturbations.

In real physical media there are usually different dissipative
mechanisms which affect wave shapes and soliton dynamics. The
influence of various types of dissipation on solitons and kinks
has been studied in detail based on various model evolution
equations (see, for instance, \cite{Grimshaw-2003}). However, the
influence of dissipation on the decay of BO solitons has not been
studied as yet. In this paper we fill this gap and study the
adiabatic decay of algebraic BO solitons under the influence of
weak dissipation of various types; Rayleigh and Reynolds
dissipation, Landau damping, dissipation in a laminar boundary
layer and Chezy friction when the soliton propagates over rough bottom topography,
as well as the dissipation caused by the radiation of small amplitude
waves in media containing weak large scale dispersion. Such
dispersion typically arises in a rotating fluid
\cite{Grimshaw-1985, Grimshaw-1998}, but can be caused by specific
dispersion dependences in other media.

If weak dissipation is taken into account the BO Eq.\ (\ref{Eq01}) is
augmented by additional terms whose structure depends on the
nature of this dissipation. In general, BO type equations with
dissipative terms can be presented in the form
\begin{equation}%
\label{Eq03}%
\frac{\partial u}{\partial t} + \alpha u \frac{\partial
u}{\partial x} + \frac{\beta}{\pi} \frac{\partial^2}{\partial
x^2}\,\wp\!\!\int\limits_{-\infty}^{+\infty}\frac{u(\xi, t)}{\xi -
x}d\xi + \delta\mathcal{D}[u] = 0,
\end{equation}
where $\mathcal{D}[u]$ is an operator which can be expressed in
the rather general form \cite{Grimshaw-2001}
\begin{equation}%
\label{Eq04}%
\mathcal{D}[u] =
\frac{1}{\sqrt{2\pi}}\int\limits_{-\infty}^{+\infty}(-ik)^m \tilde
u(k, t)e^{ikx}dk \quad \mbox{and} \quad \tilde u(k, t) =
\frac{1}{\sqrt{2\pi}}\int\limits_{-\infty}^{+\infty}u(x,
t)e^{-ikx}dx.
\end{equation}
Here $\tilde u(k, t)$ is the Fourier transform of the function
$u(x, t)$ and the parameter $m$ depends on the specific type of
dissipation.

The case $m = 0$ (together with $\delta > 0$) corresponds to
linear Rayleigh damping, as the dissipative term $\mathcal{D}[u]$
in Eq.\ (\ref{Eq03}) reduces simply to $\delta u$.  This loss has
been invoked in the internal wave context as a model for friction
in the bottom boundary layer \cite{Grimshaw-2003}. Another widely
used model in many physical contexts has $m = 2$ and $\delta < 0$.
This corresponds to Reynolds dissipation, $\mathcal{D}[u] =
-\delta u_{xx}$, and so $\delta $ represents the kinematic
viscosity of the fluid.

When $m$ is not an even number the term $(-ik)^m$
needs a more careful interpretation. In particular, when $m = 1$
it should be replaced by $|k|$.  In this case the dissipative
term $\mathcal{D}[u]$ reduces to the Hilbert transform of the
derivative $u_x$ and describes the Landau damping of plasma waves,
or of internal waves in a stratified fluid with a shear flow
\cite{Ostrovsky-1984,Ott-1969} (see Eq.\ (\ref{Eq3.1.1}) below for
an alternative representation of the BO equation with Landau
damping).

For non-integer $m$ the correct interpretation of the term $(-ik)^m$ is
\begin{equation}
\label{Eq2.102} %
(-ik)^m = |k|^m \exp{(-i\,\mbox{sign}\,[k]\,m\, \pi/2 ) }\,,
\end{equation}
where $\mbox{sign}\,[k] \equiv k/|k|$. This representation ensures
that $\mathcal{D}[u]$ is real valued when $u$ is real valued, as it
can be readily verified in this case that $\mathcal{D}[u]^{*} =
\mathcal{D}[u]$, noting that $\tilde{u}^{*}(k) = \tilde{u}(-k)$
(here the star superscript denotes the complex conjugate).
When $m = 1/2$ this dissipative term can be used for the description of wave
decay due to a laminar bottom boundary layer \cite{Grimshaw-2001}.

The operator $\mathcal{D}[u]$ can be nonlinear. An important
example is $\mathcal{D}[u] = |u|u$ for the dissipation of internal
solitary waves over a rough bottom \cite{Grimshaw-2001}. This
model is based on the empirical description of dissipation in a
turbulent boundary layer.

Note that for $m > 0$ solutions of the BO Eq.\ (\ref{Eq03})
conserve ``mass'' $M = \int\limits_{-\infty}^{\infty}u(x,t)\,dx$,
whereas for $m = 0$ the total mass is not conserved.  This issue
has been discussed by Miles \cite{Miles-1979} in the application
of the KdV equation to water waves in a channel which has variable
depth and width.  In particular, in the case of Rayleigh
dissipation, on integrating the perturbed BO Eq.\ (\ref{Eq03}) we
obtain the mass balance equation
\begin{equation}%
\label{Eq2.105}%
\frac{dM}{dt} = -\delta M\,,
\end{equation}
which has the solution $M(t) = M_0e^{-\delta t}$. It is interesting to
note that for the BO soliton (\ref{Eq02}) the total mass $M_s =
4\pi\beta/\alpha$ does not depend on its amplitude. It is then concluded that
the mass balance (\ref{Eq2.105}) implies that under the influence of even small
Rayleigh dissipation the solution $u(x, t)$ cannot be just a BO
soliton, but must contain a non-solitonic part (a trailing wave).

A shelf forms behind the evolving soliton under the influence of Chezy friction
for a similar reason based on mass balance.  In this case, the mass
balance equation gives
\begin{equation}%
\label{Eq2.106}%
\frac{dM}{dt} = -\delta \int\limits_{-\infty}^{+\infty}
|u|\,u\,dx\,.
\end{equation}
If the initial condition is a BO soliton, then the mass decay rate
at $t = 0$ is $-2\pi\beta\delta/\alpha$, which is two
times less than in the case of Rayleigh decay.  The complete
solution for $M(t)$ then requires a knowledge of the solution for $u(x,
t)$, including the non-solitonic contribution behind the soliton.

When the dissipation is sufficiently small, one would expect that
the basic structure of the solitary wave remains the same, but its
primary parameter $A$, which determines its amplitude, speed and
width, is no longer a constant, but is a slowly varying function
of time. This dependence on time can be calculated by means of
asymptotic theory \cite{Gorshkov-1981, Grimshaw-2001,
Ostrovsky-2015} which presumes that the soliton adiabatically
varies with time, while retaining its shape. This asymptotic
theory in essence reduces to an energy balance equation which
describes the time dependence of the governing soliton
parameter $A$.  In terms of a multiple scales analysis, this
energy equation is the condition for the elimination of secular terms.
Multiplying Eq.\ (\ref{Eq03}) by $u$ and then
integrating over $x$, we obtain the energy balance equation
\begin{equation}%
\label{Eq05}%
\frac{dE}{dt} = -\delta F\,,
\end{equation}
where the ``wave energy'' $E$ and dissipative function $F$ are
\begin{equation}%
\label{WaveEnerg}%
E = \frac 12 \int\limits_{-\infty}^{+\infty} u^2(x, t)\,dx\,,
\quad F = \int\limits_{-\infty}^{+\infty}u(x,
t)\,\mathcal{D}[u]\,dx\,.
\end{equation}
Using Parseval's theorem \cite{Arfken-2001}, we can calculate the
wave energy as
\begin{equation}%
\label{Eq2.103}%
E =  \frac{1}{4\pi}\int\limits^{+\infty}_{-\infty}\,|\hat{u}|^2 \,
dk\, . %
\end{equation} %
The convention (\ref{Eq2.102}) for $(-ik)^m $ ensures that $F$
is real valued.  It can then be presented in the alternative
form as
\begin{equation}%
\label{Eq2.104}%
F = \frac{\sigma}{2\pi} \int\limits^{+\infty}_{-\infty}\,
(-ik)^m\, |\tilde{u}|^2 \, dk = \frac{\sigma}{\pi}
\cos{\frac{m\pi}{2}} \int\limits^{+\infty}_{0}\,|k|^m\,
|\hat{u}|^2 \, dk\,,
\end{equation}
where $\sigma = \pm 1$. The sign $\sigma$ should be chosen so
that the dissipative function $F > 0$. In particular, for $0 \le m <
1$, $\sigma = 1$, whereas for $1 < m \le 2$, $\sigma = -1$. For $m
= 1$, that is Landau damping, expression (\ref{Eq2.104}) is invalid
(formally it gives $F = 0$).  In this case the correct expression is
\begin{equation}%
\label{Eq2.107}%
F = \frac{1}{\pi}\int\limits^{+\infty}_{0}\, |k|\, |\tilde{u}|^2
\, dk\,.
\end{equation}

The wave energy $E$ is obviously conserved when there is no
dissipation ($\delta = 0$).  For the BO soliton (\ref{Eq02})
the energy is
\begin{equation}%
\label{Eq2.109}%
E_s = \frac{\pi}{4}\Delta A^2 = \frac{\pi\beta A}{\alpha}
\end{equation}
and the soliton Fourier spectrum is $\tilde u(x) = \pi A\Delta
e^{-k/\Delta} = (4\pi\beta/\alpha)e^{-\alpha A k/4\beta}$.

Below we apply this asymptotic approach to calculate adiabatic
soliton decay under the action of the different decay mechanisms
discussed here.  We then study the non-adiabatic stage of soliton decay,
which occurs on very long time scales.

\section{Rayleigh dissipation}
\label{Sect2}%

We first consider the effect of weak Rayleigh dissipation on the
dynamics of a BO soliton.  Bearing in mind that in the case of
Rayleigh dissipation $\mathcal{D}[u] = \delta u$ (see above), we
obtain from the energy balance equation (\ref{Eq05})
\begin{equation}
\label{Eq2.1} %
\frac{dE_s}{dt} = -2\delta E_s\,, \quad E_s(t) = E_0e^{-t/\tau}\,,
\quad A(t) = A_0e^{-t/\tau}\,,
\end{equation}
where $E_0$ and $A_0$ are the initial soliton energy and
amplitude, respectively (note that $E_s \sim A$ as per Eq.\
(\ref{Eq2.109})).  $\tau = 1/2\delta$ is the characteristic
time of soliton decay (the time taken for the soliton amplitude to
decrease by the factor $e$). Note that the linear wave solution of
the linearised BO equation with $\alpha = 0$ decays two times
slower, with $\tau_{sin} = 1/\delta$. This discrepancy in the
decay rate is explained by the relationship between the soliton
amplitude and width which is absent for a linear wave. Using the
relationship between the soliton amplitude, speed and width (see
after Eq.\ (\ref{Eq02})), we find that the velocity also decreases
exponentially in time with the same decay rate, whereas the width
exponentially increases with time, $\Delta = \Delta_0 e^{t/\tau}$.

\begin{figure}[h!]
\centering %
\includegraphics[width=14cm]{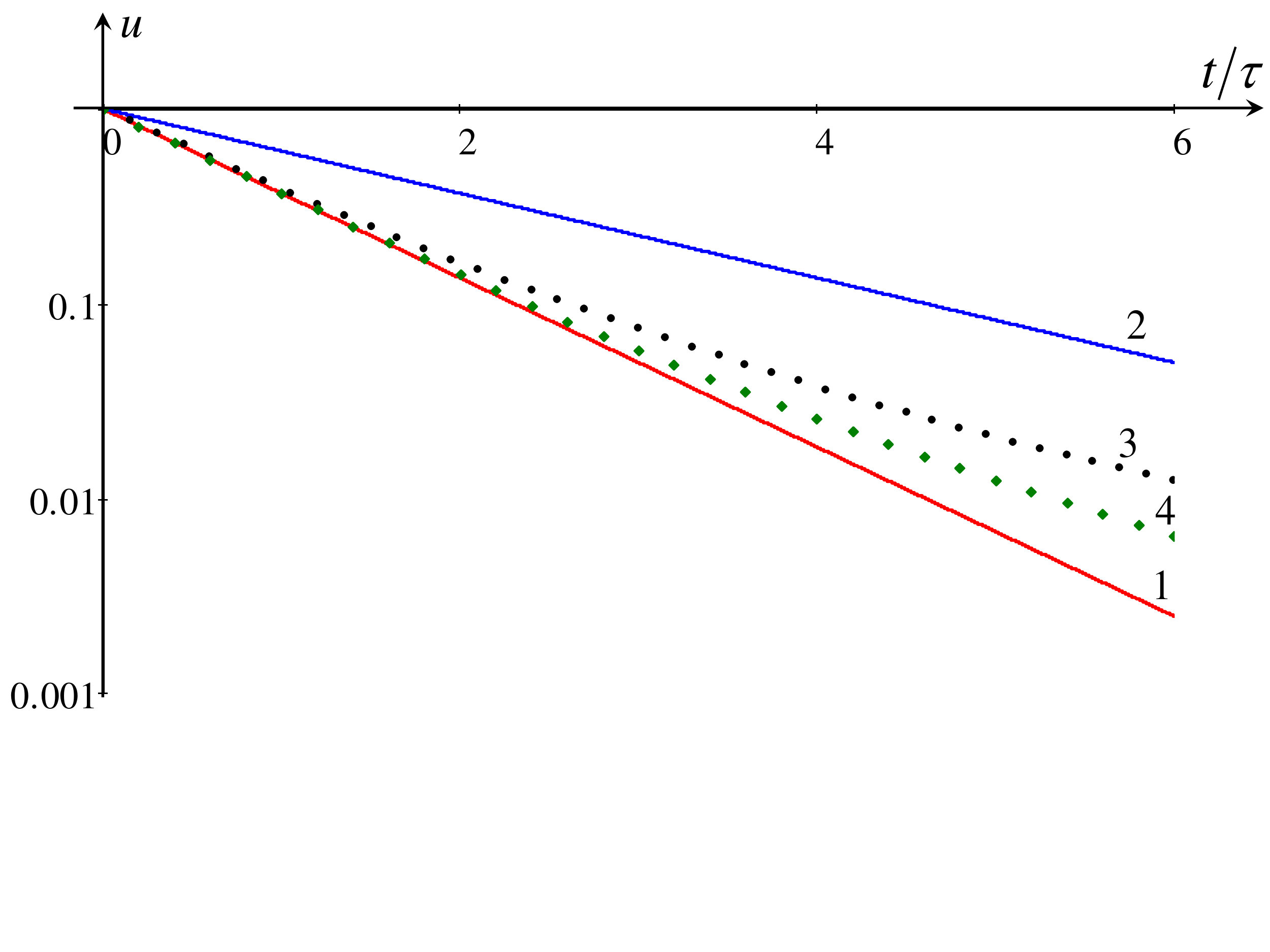}
\vspace*{-2.5cm}%
\caption{(colour online). The dependence of soliton amplitude on
time (semi-log scale) in the case of Rayleigh
dissipation. Red line 1--- the theoretical dependence; blue line 2---
the amplitude decay of a linear wave; line 3 with dots---
numerical data obtained with $\delta = 10^{-3}$; line 4 with
rhombuses--- numerical data obtained with $\delta = 10^{-4}$.
The other parameters are $\alpha = 2$ and $\beta = 1$.}%
\label{f01}
\end{figure}

To validate this asymptotic result we undertook direct numerical simulations
of soliton evolution within the framework of the perturbed BO
equation (\ref{Eq03}) with an additional dissipative term. The
dependence of $A(t)$ for Rayleigh dissipation is shown in Fig.\
\ref{f01} for two values of $\delta = 10^{-3}$ (dots) and $\delta
= 10^{-4}$ (rhombuses). As can be seen, the smaller the $\delta$,
the better the agreement between the numerical data and the
asymptotic theory (red line 1).
%%%%%%%%%%%%%%%%%%%%%%%%%%%%%%%%%%%%%%%%%%%%%%%%%%%%%%%%%%%%%%%

The range of validity of the asymptotic theory can be estimated on the
basis of different approaches. A simple estimate can be obtained
by a comparison of the relative strengths of the dissipative term ($\sim
\delta A(t)$) and the nonlinear ($\sim \alpha A^2(t)/2\Delta(t)$) or
dispersive ($\sim \beta A(t)/\Delta^2(t)$) terms (it is worth
recalling that there is a balance of nonlinear and dispersive effects
for a soliton solution, therefore any of these latter terms can be
taken for the estimate). In particular, the ratio of dissipative
to nonlinear terms is $8\delta\beta/\alpha^2 A^2(t) = 8\delta\beta
e^{4\delta t}/\alpha^2 A_0^2$. This shows that the relative role
of the dissipative term quickly increases.  The adiabatic theory then
becomes invalid when this ratio becomes of order of unity.
We then obtain that the asymptotic theory is valid until
\begin{equation}
\label{Eq2.101} %
t/\tau \sim \ln{\frac{\alpha A_0}{2\sqrt{\beta\delta}}}.
\end{equation}
Using the parameters for Fig.\ \ref{f01}, we obtain $t/\tau \approx 3.1$ for
$\delta = 10^{-3}$ and $t/\tau \approx 4.3$ for $\delta = 10^{-4}$, which is
in agreement with the results shown in the Figure.

Another estimate for the range of validity of the asymptotic theory can be obtained from
the mass balance equation (\ref{Eq2.105}).  Assuming that in the course of the wave evolution a
shelf $u_{shelf}$ is generated behind the leading solitary wave, the total wave mass is
\begin{equation}%
\label{raymass1}%
M =  \int\limits_{-\infty}^{P}\, u_{shelf} \, dx + M_{s}\,. %
\end{equation}
Here $P(t)$ is solitary wave position, such that $P_t = V(t) =
V_0e^{-2\delta t}$.  Differentiating $M$ with respect to $t$ and using the mass balance
equation (\ref{Eq2.105}), we obtain
  \begin{equation}%
  \label{raytail}%
V(t)u_{shelf} + \frac{\partial M_{s}}{\partial t} + \delta
\left(M_{s} + M_{shelf}\right) = 0\,. %
\end{equation}
Now the soliton mass $M_{s} = 4\pi\beta/\alpha$ is
independent of time (see above).  Therefore, neglecting $M_{shelf}$
in comparison with $M_{s}$, we obtain to first order in the parameter
$\delta$
\begin{equation}%
\label{raytail1}%
u_{shelf} = -\frac{\delta M_{s}}{V(t)}\,.
\end{equation}
Substituting the expressions for the soliton mass $M_s$ and
speed $V(t)$, we obtain
\begin{equation}
u_{shelf} = -16\pi \delta \beta e^{2\delta t}/\alpha^2A_0.
\label{e:raymass}
\end{equation}
The adiabatic theory breaks down when $u_{s} \sim u_{shelf}$, which then yields
\begin{equation}%
\label{raytail2}%
\frac{16\pi \beta \delta }{A_{0}\alpha^2}e^{2\delta t} \sim A_0e^{-2\delta t}\,.%
\end{equation}
This can happen quite quickly, especially for a large initial
amplitude $A_0$, and the adiabatic theory breaks down
when $t/\tau \sim \log{\left(\alpha
A_0/4\sqrt{\pi\beta\delta}\right)}$. This estimate is similar to
that derived above, Eq.\ (\ref{Eq2.101}), up to a numerical
constant.  However, it agrees a bit better with the numerical data
shown in Fig.\ \ref{f01}.  In particular, it gives that the asymptotic theory
breaks down when $t/\tau \approx 2.2$ for $\delta = 10^{-3}$ and $t/\tau \approx 3.3$ for $\delta =
10^{-4}$.

These estimates explain the deviation of the numerical data from the
theoretical line after a certain time, as seen in Fig.\ \ref{f01}. As has
been noted by Miles \cite{Miles-1979}, the results obtained within
the framework of asymptotic theory when the total ``wave mass'' $M
\ne 0$ ``must be regarded with some caution. It appears likely
that such solutions cannot be uniformly valid as $t \to \infty$,
but this does not exclude the possibility that they are viable
approximations in some contexts.''

When a shelf forms behind the soliton it obeys long wave
dynamics initially, so that it is governed by the approximate equation
\be%
\label{rayshelf}%
\frac{\partial u_{shelf}}{\partial t} + \delta u_{shelf} \approx 0
\,, \quad \mbox{and} \quad u_{shelf} \approx
u_{shelf}\left(x=P\right)\exp{\left\{-\delta\left[t -
R(x)\right]\right\}} \,,
\ee%
where $x=P(t)$ defines $t = R(x)$. Thus the shelf, which lies in $x <
P$, decays exponentially behind the soliton.
%%%%%%%%%%%%%%%%%%%%%%%%%%%%%%%%%%%%%%%%%%%%%%%%%%%%%%%%%%%%%%%
\begin{figure}[t!]
\centering %
\includegraphics[width=14.0cm]{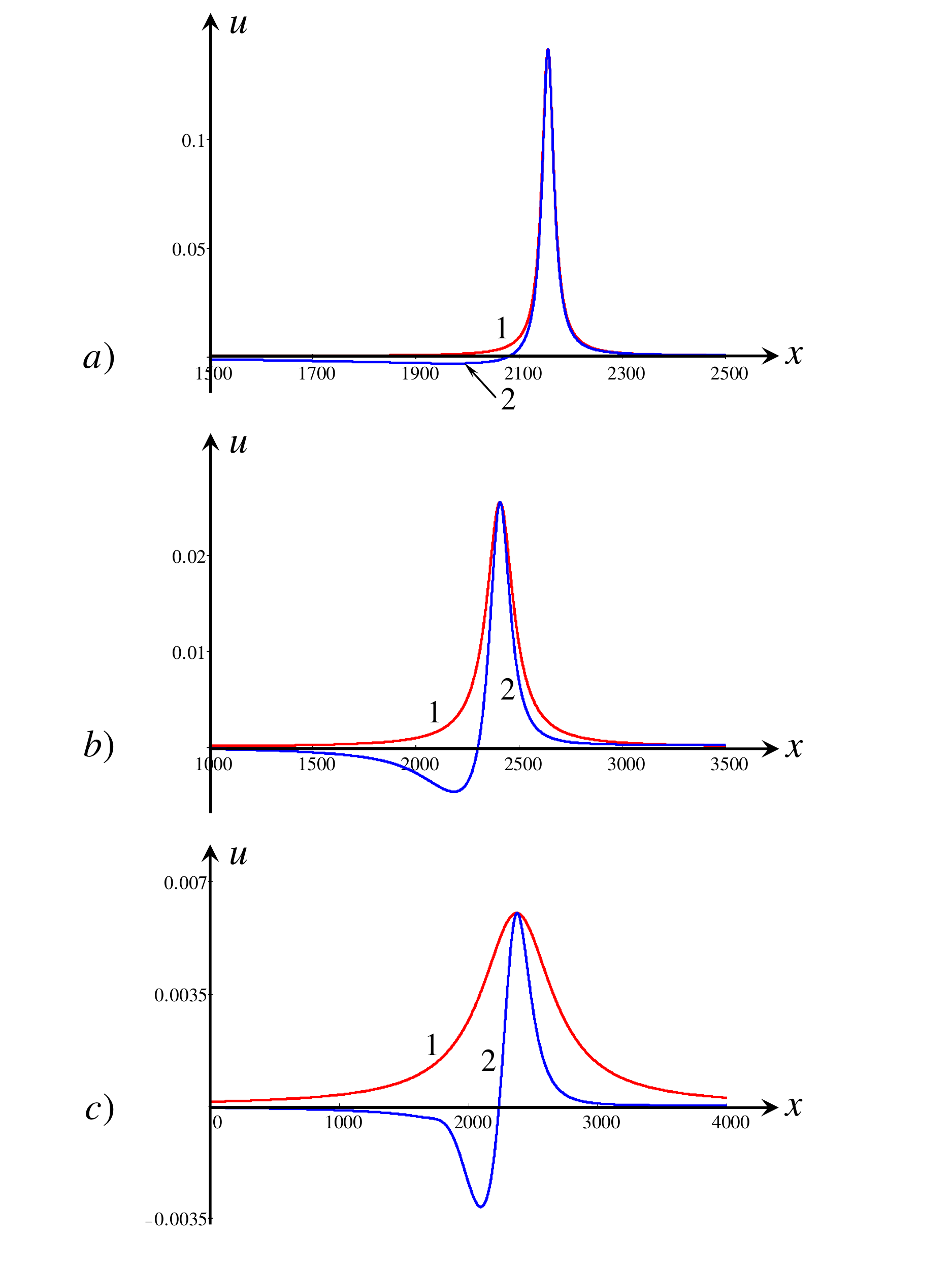}
\vspace*{-0.5cm}%
\caption{(colour online). Soliton decay due to Rayleigh
dissipation. Solitary wave profiles (lines 2) at different
instants of time compared with the BO soliton of the same
amplitudes (lines 1): a) $t = 10,000$ ($t/\tau = 2$); b) $t =
20,000$ ($t/\tau = 4$); c) $t = 30,000$ ($t/\tau =
6$). Here $\alpha = 2$, $\beta = 1$ and $\delta = 10^{-4}$.} %
\label{f02}
\end{figure}

%\clearpage

Due to the influence of dissipation, the profile of the leading
pulse deviates from a soliton profile. Figure \ref{f02}
illustrates this phenomenon for the particular instants of time $t
= 10,000$ (frame a) and $t = 20,000$ (frame b). As can be seen,
the rear part of the leading pulse significantly differs from the
soliton profile and even has an opposite polarity (see lines 2).
At $t = 20,000$ even the frontal part differs from the soliton
profile. The asymptotic formula for the soliton amplitude decay
(\ref{Eq2.1}) assumes that energy is conserved by the soliton.
However, it must lose a small amount of energy due to dissipation.
This could be accounted for in the higher orders of the asymptotic
expansion. For large time, the tail generated behind the soliton
gradually accumulates a non-negligible amount of energy (see Fig.
\ref{f02}), which is not accounted for in the first order
asymptotic theory.  After this time, expression (\ref{Eq2.1}) for
the soliton decay ceases to be valid.

\section{Reynolds dissipation}
\label{Sect3}%

We shall now consider the effect of weak Reynolds dissipation on
the dynamics of a BO soliton. The BO equation with Reynolds
dissipation is Eq.\ (\ref{Eq03}) with the dissipative term
$\mathcal{D}[u]=-\delta u_{xx}$ instead of $\delta u$, where the
indices of $u$ represent derivatives with respect to $x$. The
energy balance equation (\ref{Eq05}) then gives
\begin{equation}%
\label{Eq08}%
\frac{dE}{dt} = -\delta
\int\limits_{-\infty}^{+\infty}\left(\frac{\partial u}{\partial
x}\right)^2\,dx.
\end{equation}
The wave energy $E$ now decays non-exponentially.
Substituting the soliton solution (\ref{Eq02}) into the energy evolution Eq.\ (\ref{Eq08})
and integrating, we obtain the soliton amplitude evolution
\begin{equation}%
\label{Eq09}%
\frac{dA}{dt} = -\frac{\delta\alpha^2}{16\beta^2}A^3.
\end{equation}
After integration of this equation we obtain
\begin{equation}%
\label{Eq10}%
A(t) = \frac{A_0}{\sqrt{1 + t/\tau}}, \quad \tau =
\frac{8\beta^2}{\delta \alpha^2A_0^2} =
\frac{\Delta_0^2}{2\delta}.
\end{equation}

%%%%%%%%%%%%%%%%%%%%%%%%%%%%%%%%%%%%%%%%%%%%%%%%%%%%%%%%%%%%%%%
\begin{figure}[t!]
\centering %
\includegraphics[width=14cm]{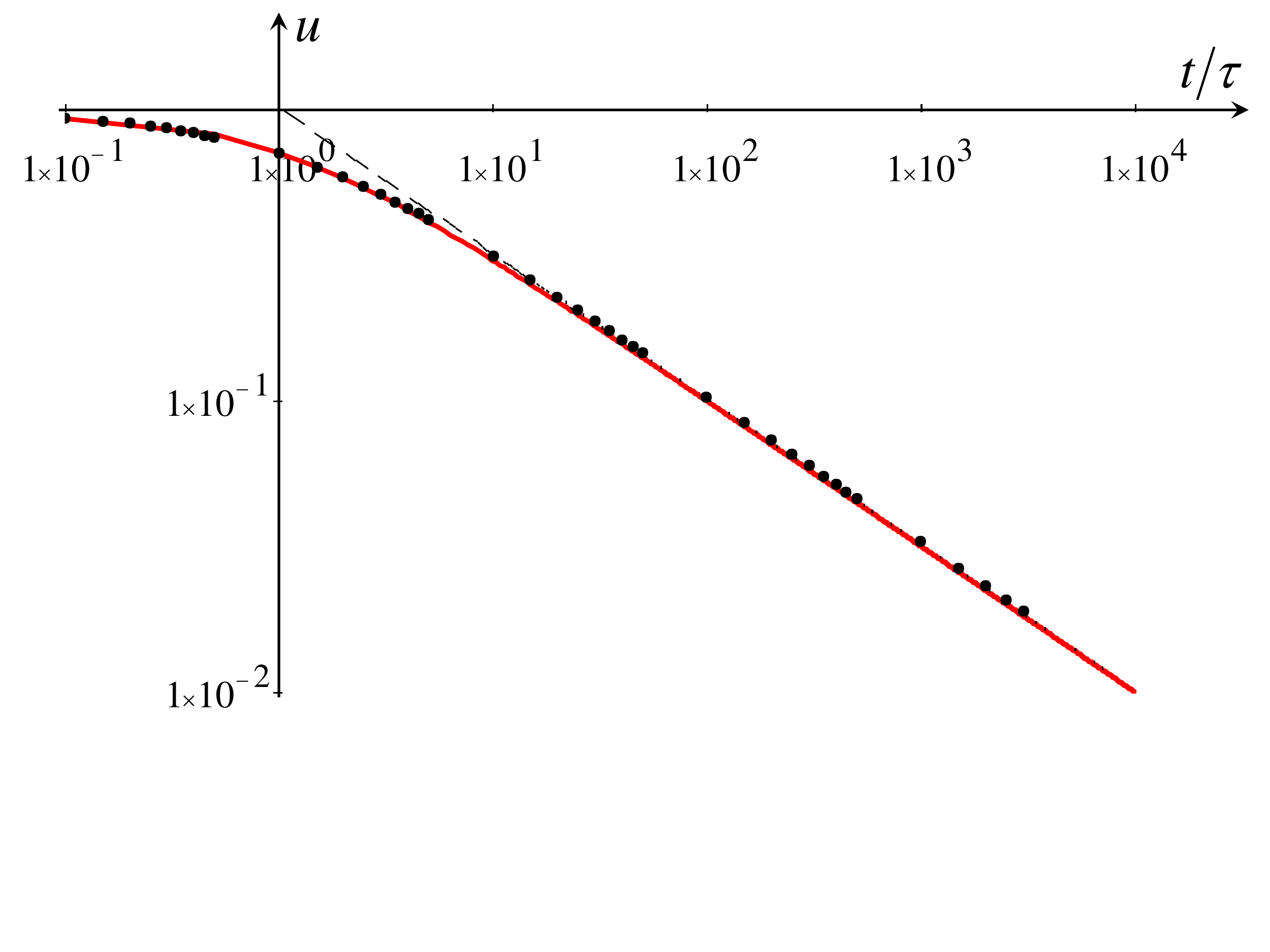}
\vspace*{-2.5cm}%
\caption{(colour online). Soliton amplitude dependence (log-log
scale) on the normalised time $t/\tau$ for Reynolds dissipation.
Solid line --- theoretical dependence with $\delta = 0.1$; dots
--- numerical data for $\alpha = 2$ and $\beta = 1$; dashed line
--- asymptotic dependence $u
\sim (t/\tau)^{-1/2}$.} %
\label{f03}
\end{figure}

The theoretical dependence of the amplitude $A(t)$ is depicted in
Fig.\ \ref{f03}, together with the numerical solution. Note that
for this type of dissipation the agreement between the numerical
data and adiabatic theory is very good, even for a relatively large
dissipation coefficient, $\delta = 0.1$. This can again be explained
with the help of simple estimates of the relative strength of
the dissipative term ($\sim \delta A(t)/\Delta^2(t)$) in comparison
with the nonlinear or dispersive terms (see Sect.\ \ref{Sect2}).
The ratio on the dissipative term with respect to the nonlinear
term is $\delta/2\beta$; it does not depend on time and remains
small if it is initially small.  It is also worth recalling that
for Reynolds dissipation the total wave mass is conserved, based on the
perturbed BO Eq.\ (\ref{Eq03}).

\begin{figure}[h]
\centering %
%\vspace*{2.0cm}%
\includegraphics[width=12cm]{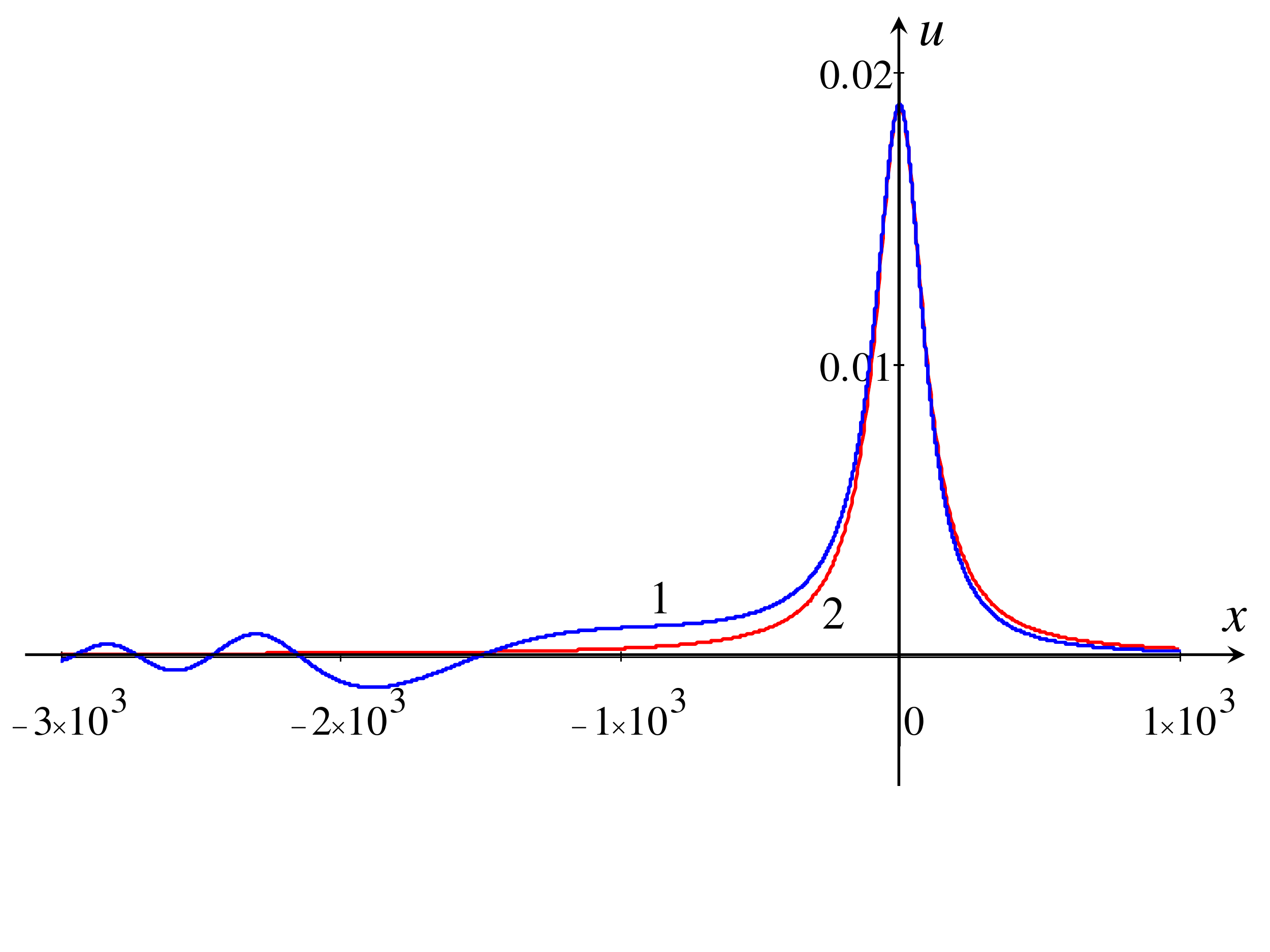}
\vspace*{-1.5cm}%
\caption{(colour online). Solitary wave profile at $t = 60,000$
($t/\tau = 3,000$) (line 1) for the BO equation (\ref{Eq03}) with
Reynolds dissipation. Line 2 represents a BO soliton of the
same amplitude as the leading pulse shown by line 1. The
parameters are $\alpha = 2$, $\beta = 1$ and $\delta = 0.1$.}%
\label{f04}
\end{figure}

A linear wave decays exponentially due to Reynolds dissipation,
$u_{lin} \sim \exp{\left(-\delta k^2 t\right)}$, where $k$ is the
wavenumber. Figure \ref{f04} illustrates the solitary wave profile
at $t = 60,000$ (line 1).  At this large time the wave profile
deviates slightly from a BO soliton of the same amplitude (line
1). One can clearly see the formation of a shelf behind the
leading pulse in the near field, whereas in the far field the
numerical solution is a decaying periodic wave.

\section{Landau damping}
\label{Sect4}%

This type of soliton decay applies particularly to
nonlinear waves in plasma \cite{Ott-1969}.  The decay of plane solitons
within the framework of the KdV equation augmented by a dissipative
term describing Landau damping has been studied as early as
1969 \cite{Ott-1969}. In the case of the BO equation (\ref{Eq03}),
Landau damping corresponds to the integro-differential
operator $\mathcal{D}[u]$ in Eq.\ (\ref{Eq04}) with the index $m =
1$ in the spectral operator.  The corresponding BO equation can be presented in explicit form as
\begin{equation}%
\label{Eq3.1.1}%
\frac{\partial u}{\partial t} + \alpha u \frac{\partial
u}{\partial x} + \frac{\beta}{\pi} \frac{\partial^2}{\partial
x^2}\,\wp\!\!\int\limits_{-\infty}^{+\infty}\frac{u(\xi, t)}{\xi -
x}d\xi - \frac{\delta}{\pi} \frac{\partial}{\partial
x}\,\wp\!\!\int\limits_{-\infty}^{+\infty}\frac{u(\xi, t)}{\xi -
x}d\xi = 0.
\end{equation}

The energy balance equation (\ref{Eq05}) determining the evolution of the soliton
parameters is then
\begin{equation}%
\label{Eq3.1.2}%
\frac{dE}{dt} = \frac{\delta}{\pi}\left\langle
u(x,t)\frac{\partial}{\partial
x}\,\wp\!\!\int\limits_{-\infty}^{+\infty}\frac{u(\xi, t)}{\xi -
x}d\xi\right\rangle,
\end{equation}
where the angular brackets stand for integration in $x$ over the
infinite domain.  The expression in the right hand side can be directly evaluated,
yielding
\begin{equation}%
\label{Eq3.1.3}%
\frac{\delta}{\pi}\left\langle u(x,t)\frac{\partial}{\partial
x}\,\wp\!\!\int\limits_{-\infty}^{+\infty}\frac{u(\xi, t)}{\xi -
x}d\xi\right\rangle = -\frac{\pi}{4}\delta A^2.
\end{equation}
Then using the expression for the soliton energy $E_s$
(\ref{Eq2.109}), we obtain
\begin{equation}%
\label{Eq3.1.4}%
\frac{dA}{dt} = -\frac{\delta\alpha A^2}{4\beta},
\end{equation}
whose solution is
\begin{equation}%
\label{Eq3.1.5}%
A(t) = \frac{A_0}{1 + t/\tau}, \quad \tau = \frac{4\beta}{\delta
\alpha A_0} = \frac{\Delta_0}{\delta}.
\end{equation}

\begin{figure}[h!]
\centering %
\includegraphics[width=14cm]{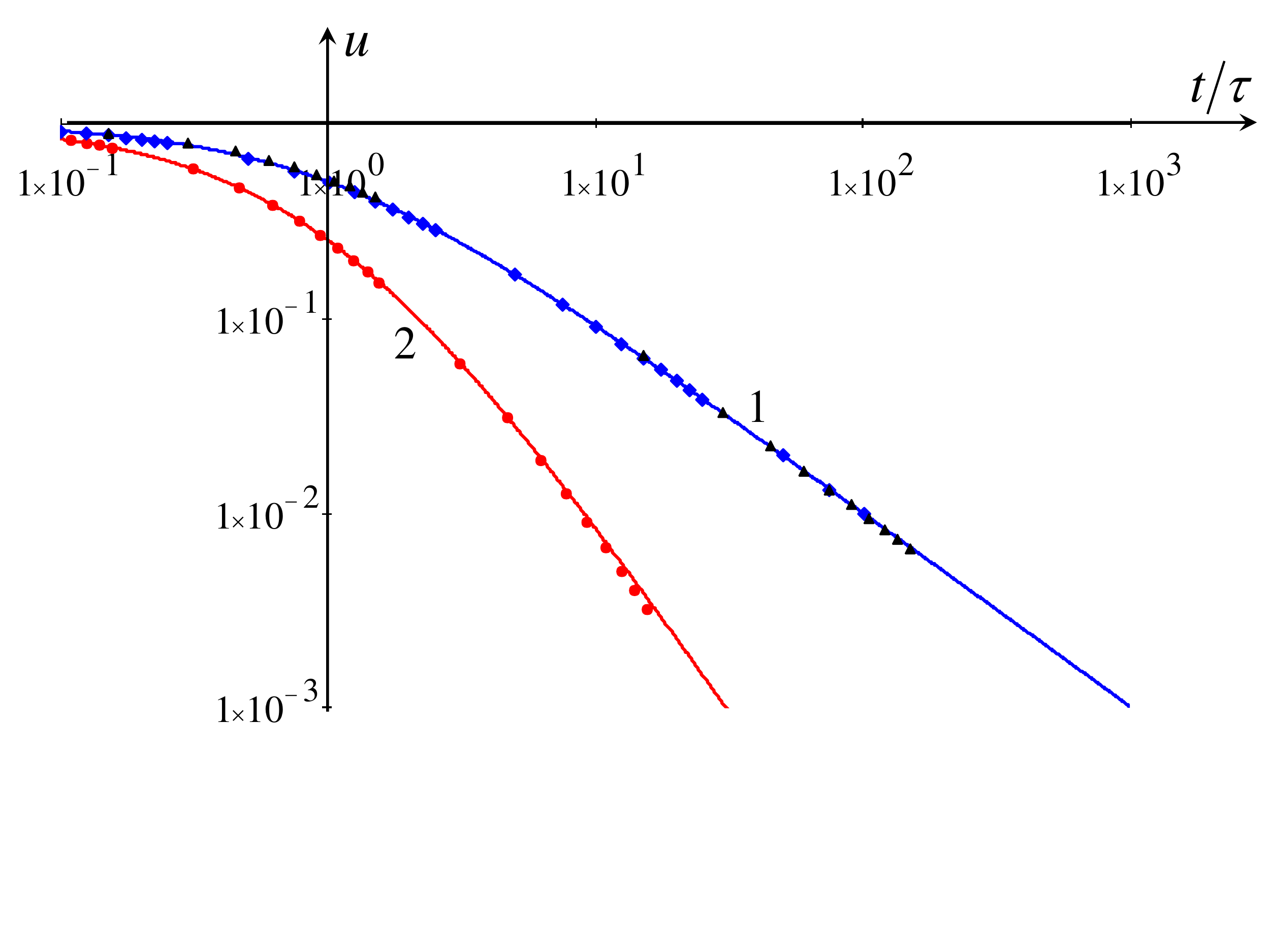}
\vspace*{-2.5cm}%
\caption{(colour online). Soliton amplitude dependences (log-log
scale) on the normalised time $t/\tau$ in the cases of Landau
damping ($\delta = 0.01$) and Chezy friction ($\delta = 0.01$)
(line 1), and dissipation in a laminar boundary layer ($\delta =
0.001$) (line 2). Dashed lines show the asymptotic dependences at
large time ($t \to \infty$): $u \sim (t/\tau)^{-1}$ for line 1 and
$u \sim (t/\tau)^{-2}$ for line 2, respectively. Symbols show the
numerical data; rhombuses for Landau damping and triangles for
Chezy friction (not all data is
shown to avoid overlapping of symbols).  The other parameters are $\alpha = 2$ and $\beta = 1$.} %
\label{f05}
\end{figure}

\begin{figure}[h!]
\centering %
\includegraphics[width=12cm]{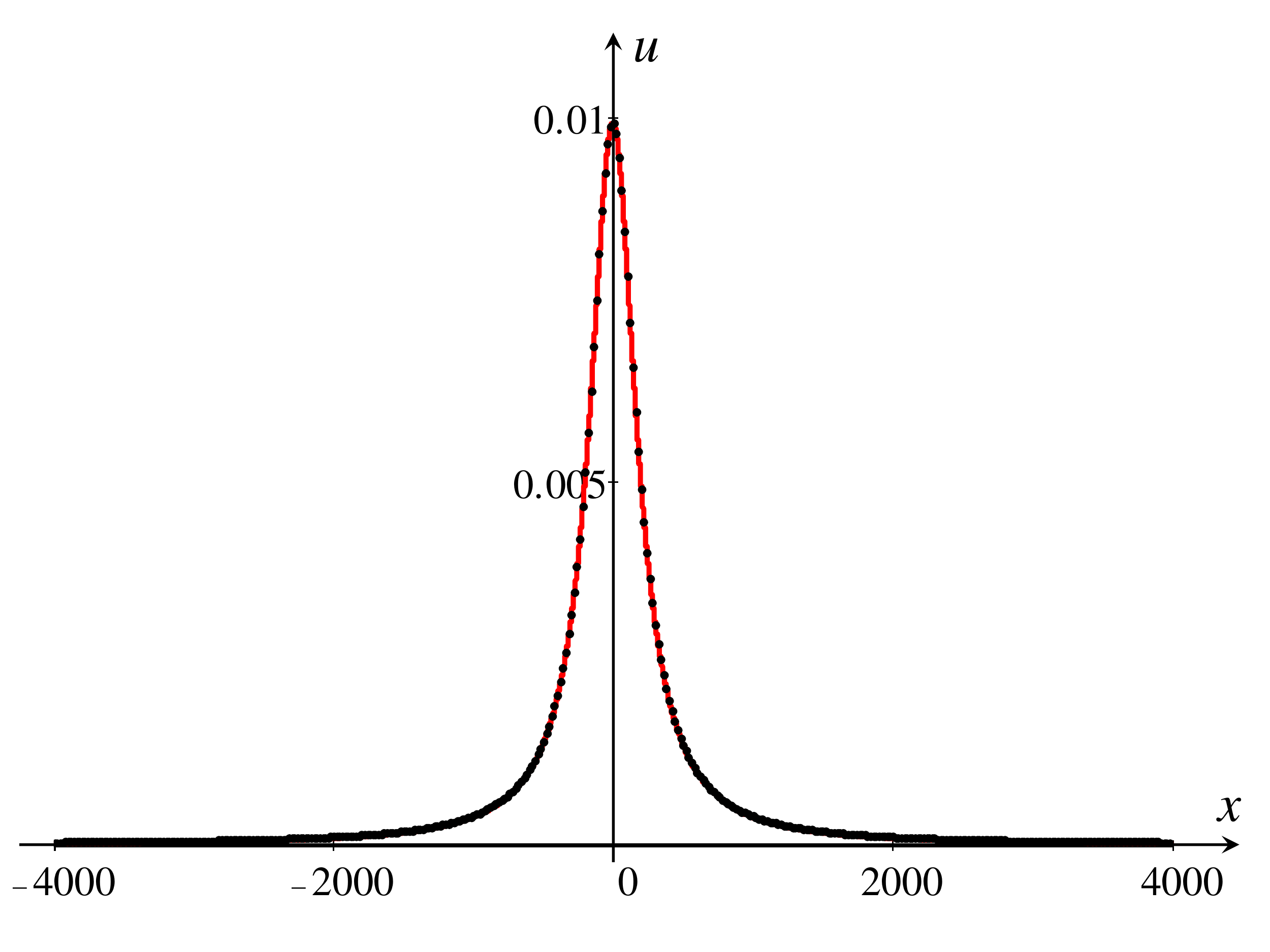}
\vspace*{-0.5cm}%
\caption{(colour online). Solitary wave profile at $t = 20,000$
($t/\tau = 100$) in the case of Landau damping. Dots show the
numerical data for $\alpha = 2$, $\beta = 1$ and $\delta = 0.01$.}%
\label{f06}
\end{figure}

The asymptotic dependence of the soliton amplitude on the normalised time is
shown in Fig.\ \ref{f05} by line 1. A simple estimate of the
relative strength of the dissipative term ($\sim \delta
A(t)/\Delta(t)$) to the dispersive or nonlinear
terms (see Sect.\ \ref{Sect2}) gives $\delta\Delta(t)/\beta \sim
\delta(1 + t/\tau)/\alpha A_0$. This ratio gradually increases with
time, so that the adiabatic theory should become less valid as time increases.
However, the solitary wave profile exactly coincides with the profile of a BO
soliton of the same amplitude for all times and no shelf behind the leading
solitary wave is generated (see, e.g., Fig.\ \ref{f06}). Such good
agreement between the numerical data and asymptotic theory
suggests that a soliton solution of the form
%%%%%%%%%%%%%%%%%%%%%%%%%%%%%%%%%%%%%%%%%%%%%%%%%%%%%%%%%%%%%%%

%%%%%%%%%%%%%%%%%%%%%%%%%%%%%%%%%%%%%%%%%%%%%%%%%%%%%%%%%%%%%%%

\begin{equation}%
\label{Eq3.1.6}%
u(x, t) = \frac{A(t)}{1 + \xi^2/\Delta^2(t)}, \quad \mbox{where}
\quad \xi = x - \int V(t)\,dt\,
\end{equation}
can be an exact solution of Eq.\ (\ref{Eq3.1.1}), provided that
$A(t)$ is given by Eq.\ (\ref{Eq3.1.5}), $\Delta = 4\beta/\alpha
A(t)$ and $V(t) = \alpha A(t)/4$. Direct substitution of this
solution into Eq.\ (\ref{Eq3.1.1}) confirms that this is the case. This
is a nice example of an exact non-stationary solution in the form of a
decaying soliton of a nonlinear equation with dissipation.
It is worth recalling that for Landau damping the total wave
mass is conserved, based on Eq.\ (\ref{Eq03}). As the soliton mass is a
constant, then formally no trailing shelf can be formed at
leading order. This implies that the adiabatic theory will hold for
large time.  In fact, it is valid for all time in the case of
Landau damping.

A linear wave governed by the damped BO Eq.\ (\ref{Eq03}) decays exponentially
$u_{lin} \sim \exp{\left(-\delta k t\right)}$, where $k$ is the
wavenumber.  For a KdV soliton, the amplitude decay due to Landau damping is
more rapid \cite{Ott-1969, Ott-1970}, $A(t) = A_0\left(1 +
t/\tau\right)^{-2}$, due to the specific relationship between the
soliton width $\Delta$ and amplitude $A$.

\section{Soliton decay due to dissipation in a laminar boundary layer}
\label{Sect5}%

In some flows a pycnocline (a sharp density interface in a
stratified fluid) may be located near a boundary, so that the
near-bottom layer is relatively thin and overlayed by a very deep
upper layer. Long (in comparison with the thickness of the lower
layer) internal waves propagating on a pycnocline are described by
the BO equation. In these situations, wave dissipation can be caused
by a laminar boundary layer or wave scattering on random bottom
irregularities. In the former case the model equation is the BO
equation (\ref{Eq03}) with the linear operator $\mathcal{D}[u]$
(\ref{Eq04}) with $m = 1/2$, whereas in the latter case the
appropriate model is equation (\ref{Eq03}) with nonlinear
Chezy friction described by the operator $\mathcal{D}[u] = |u|u$.
In this section we shall study soliton decay caused by dissipation
in a laminar boundary layer, and in the next section decay
due to Chezy friction.

The dissipative function $F$ in the case of dissipation in a
laminar boundary layer can be calculated with the help of Eq.\
(\ref{Eq2.104}) with $m = 1/2$, to yield
\begin{equation}%
\label{Eq3.2.3}%
F = \frac{8\sqrt{2}}{\pi}\int\limits^{+\infty}_{0}\sqrt{k}\,
\left(\frac{\pi\beta}{\alpha}\right)^2e^{-2k\Delta}\, dk\, =
\frac{\pi}{4}\sqrt{\frac{\pi\beta}{\alpha}}A^{3/2}.
\end{equation}
Then, using the energy balance equation (\ref{Eq05}) and expression (\ref{Eq2.109})
for the soliton energy, we can calculate the dependence of the soliton amplitude on time
\begin{equation}%
\label{Eq3.2.5}%
A(t) = \frac{A_0}{\left(1 + t/\tau\right)^2}, \quad \mbox{where}
\quad \tau = \frac{8}{\delta}\sqrt{\frac{\beta}{\pi\alpha A_0}} =
\frac{4}{\delta}\sqrt{\frac{\Delta_0}{\pi}}.
\end{equation}
The dependence of the soliton amplitude on the normalised time is
shown in Fig.\ \ref{f05} by line 2. The total wave mass is conserved
within Eq.\ (\ref{Eq03}) with the appropriate decay term. A linear wave in this
case decays exponentially, $u_{lin} \sim \exp{\left(-\delta
\sqrt{k} t\right)}$, where $k$ is the wavenumber.
A simple estimate of the relative strength of the dissipative term
($\sim \delta A(t)/\Delta^{3/2}(t)$) in comparison with the
nonlinear and dispersive terms (see Sect.\ \ref{Sect2}) shows that
the ratio of these terms is $\delta/\sqrt{\alpha\beta A(t)} =
\delta (1 + t/\tau)/\sqrt{\alpha \beta A_0}$, which increases with
time. This means that the asymptotic theory ceases to be valid when this
ratio is unity, i.e., when $t/\tau \approx \sqrt{\alpha \beta A_0}/\delta$. For $\delta
= 10^{-3}$, $\alpha = 2$ and $\beta = A_0 = 1$, this gives $t/\tau
\approx \sqrt{2}\times 10^3$.
%%%%%%%%%%%%%%%%%%%%%%%%%%%%%%%%%%%%%%%%%%%%%%%%%%%%%%%%%%%%%%%
\vspace*{-0.5cm}%
\begin{figure}[h!]
\centering %
\includegraphics[width=14.0cm]{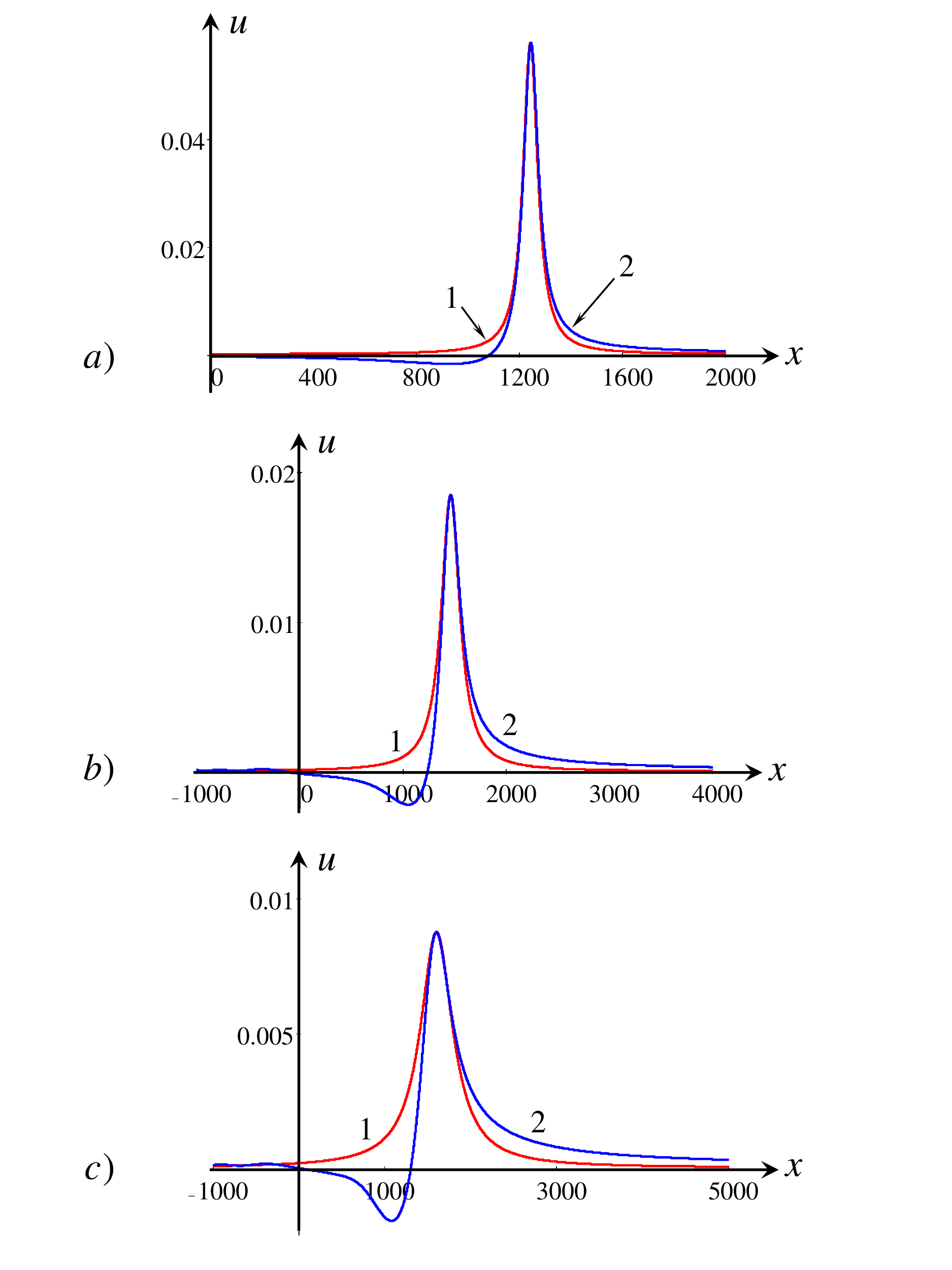}
\vspace*{-0.5cm}%
\caption{(colour online). Soliton damping due to dissipation in a laminar boundary
layer. Solitary wave profiles (lines 2) at different instants of
time compared with BO solitons of the same amplitudes
(lines 1): a) $t = 10,000$ ($t/\tau = 3.133$); b) $t =
20,000$ ($t/\tau = 6.267$); c) $t = 30,000$ ($t/\tau = 9.4$).
Other parameters are $\alpha = 2$, $\beta = 1$ and $\delta = 0.001$.} %
\label{f07}
\end{figure}

Figure \ref{f07} illustrates the deformation of the solitary wave
profile in the course of dissipation.  In this case mass is
conserved too. Therefore, no trailing shelf can be formed at
leading order. This again implies that the adiabatic theory will
hold for a long time.  However, solution (\ref{Eq02}) with the
amplitude decaying as per Eq.\ (\ref{Eq3.2.5}) is not an exact
solution in this case.

%\clearpage

For a KdV soliton the amplitude decay due to dissipation in a
laminar boundary layer is more rapid \cite{Grimshaw-2001}, $A(t) =
A_0\left(1 + t/\tau\right)^{-4}$, due to the specific relationship
between the soliton width $\Delta$ and amplitude $A$.

\section{Soliton decay due to Chezy friction}
\label{Sect6}%

\begin{figure}[h!]
\centering %
\includegraphics[width=14.0cm]{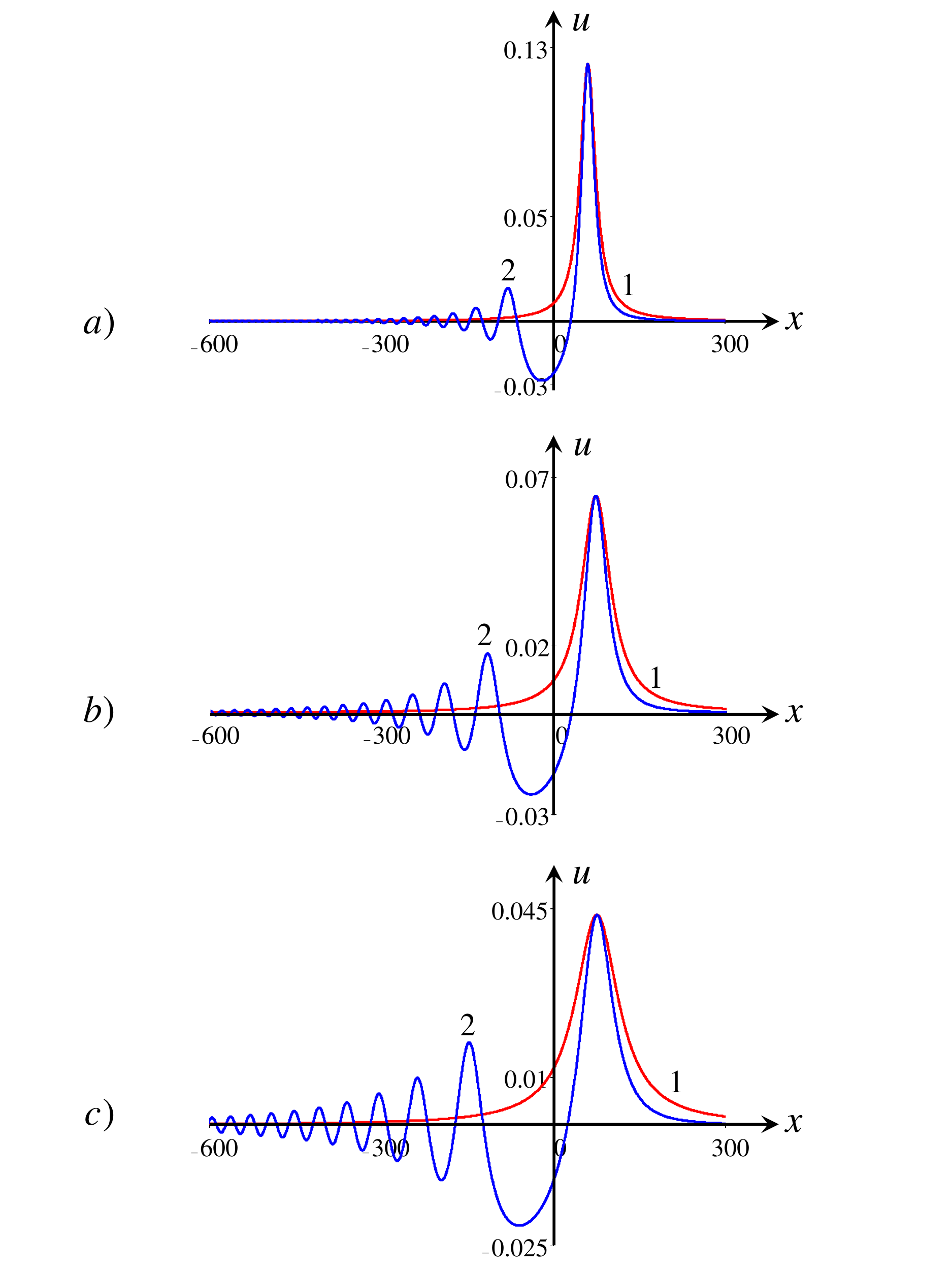}
\vspace*{-0.5cm}%
\caption{(colour online). Soliton damping due to Chezy friction.
Wave profiles (lines 2) at different instants of time compared
with BO solitons of the same amplitudes (lines 1):
a) $t = 500$ ($t/\tau = 7.5$); b) $t = 1000$ ($t/\tau = 15$;
c) $t = 1,500$ ($t/\tau = 22.5$).
The other parameters are $\alpha = 2$, $\beta = 1$ and $\delta = 0.01$.} %
\label{f08}
\end{figure}

As was mentioned above, in the case of nonlinear waves
propagating on a pycnocline located near a bottom in a deep fluid
(for example, in the two layer model of the atmosphere or ocean)
Chezy friction can be important. In general, Chezy
friction is used as an empirical model to describe wave energy
losses due to scattering on random bottom roughness (see, e.g.,
Ref.\ \cite{Internet}). In this case the damped BO Eq.\ (\ref{Eq03}) contains the
dissipation coefficient $\delta |u|$ which is not a constant, but
depends on the modulus of the wave field, and the value of
$\delta$ depends on the degree of roughness of the bottom
\cite{Grimshaw-2001}. The energy balance equation is now
\begin{equation}%
\label{Eq3.3.1}%
\frac{dE}{dt} = -\delta \int\limits_{-\infty}^{+\infty}|u|u^2\,dx
= -\frac{3\pi\delta}{8}A^3\Delta =
-\frac{3\pi\beta\delta}{2\alpha}A^2.
\end{equation}
Here, we have substituted the expression for $u$ in the form of a BO
soliton (\ref{Eq03}) and used the relationship between its width
$\Delta$ and amplitude $A$.  From this energy equation we obtain the soliton amplitude
evolution
\begin{equation}%
\label{Eq3.3.2}%
\frac{dA}{dt} = -\frac 32\delta A^2 \quad \mbox{and} \quad A(t) =
\frac{A_0}{1 + t/\tau},
\end{equation}
where $\tau = 2/(3\delta A_0)$.
The dependence of the soliton amplitude on the normalised time is the
same as in the case of Landau damping (cf.\ Eq.\ (\ref{Eq3.1.5})) and
is shown in Fig.\ \ref{f05} by line 1. In the same figure we
present the numerical data (shown by triangles), which is in a
good agreement with the theoretical prediction. In Fig.\ \ref{f08}
the deformation of soliton profile at different instances of time is shown.
%%%%%%%%%%%%%%%%%%%%%%%%%%%%%%%%%%%%%%%%%%%%%%%%%%%%%%%%%%%%%%%

The range of validity of the adiabatic theory can be estimated
again either by comparison of the relative strength of the dissipative
term ($\sim \delta A^2(t)$) to the nonlinear and dispersive
terms (see Sect.\ \ref{Sect2}), or with the help of the mass balance
equation (\ref{Eq2.105}).
In the former case, if we estimate the ratio of the dissipative to
the nonlinear terms, we obtain $t/\tau \approx \alpha^2
A_0/8\beta\delta$ (in dimensional variables $t \approx
\alpha^2/12\beta\delta^2$ is independent of the initial soliton
amplitude). Using the same parameters as above, we obtain $t/\tau
\approx 50$. However, surprisingly the numerical data demonstrates
very good agreement with the asymptotic theory, even for $t/\tau =
300$ (see the triangles in Fig.\ \ref{f05}), whereas the solitary wave
profile notably differs from the profile of a BO soliton (see Fig.\
\ref{f08}).

As mentioned, another estimate of the range of validity of the asymptotic theory
follows from the mass balance equation, which can be presented in the form
\begin{equation}%
\label{Eq3.3.3}%
\frac{dM_{shelf}}{dt} + \frac{dM_{s}}{dt} = -\delta
\int\limits_{-\infty}^{+\infty}|u|\,u\,dx = -2\delta E_s\,.
\end{equation}
Bearing in mind again that $dM_s/dt = 0$ for a BO soliton and
$dM_{self}/dt = u_{shelf}V(t)$, we obtain (cf.\ Eq.\ (\ref{raytail1}))
\begin{equation}%
\label{Eq3.3.4}%
u_{shelf} = -2\delta \frac{E_s}{V(t)} =  -8\pi\,
\frac{\beta\delta}{\alpha^2}\,.
\end{equation}
Remarkably the shelf is independent of the solitary wave amplitude
and is a constant. The adiabatic theory breaks down when the
soliton amplitude $A(t)$ decreases to this value.  Then using
Eq.\ (\ref{Eq3.3.2}), we finally obtain $t/\tau \approx
\alpha^2A_0/8\pi\beta\delta$ as an estimate of when the asymptotic
theory ceases to be valid. This gives, however, a slightly worse
estimate for the limit of validity of the asymptotic theory compared with
that obtained above.  This is due to the factor $\pi$ in the denominator.
In any case, both estimates show that the breakdown of the adiabatic
theory is rather slow in the case of Chezy decay.

When the shelf forms it obeys long wave dynamics initially, so that it
obeys the approximate equation
\be%
\label{cheshelf} %
\frac{\partial u_{shelf}}{\partial t} + \delta |u_{shelf}|
u_{shelf} \approx 0 \quad \mbox{and} \quad u_{shelf} \approx
-\frac{1}{t + F(x)} \,, %
\ee %
where $x=P(t)$ defines $t = R(x)$. The arbitrary function
$F(x)$ can be evaluated by requiring that $u_{shelf}(t - R(x))$ is
given by Eq.\ (\ref{Eq3.3.4}) at $t = R(x)$. That is,
\be%
\label{detF} %
\frac{18\pi \beta \delta }{\alpha^2 } = \frac{1}{R(x) + F(x)} \,,
\quad \quad u_{shelf} \approx - \frac{1}{t -R(x)
+ \alpha^2 /(8 \pi \beta \delta )} \,.%
\ee
Thus, the shelf again exists in $x < P$, but now decays only
algebraically behind the solitary wave (cf.\ Eq.\ (\ref{rayshelf})).

Note also that the character of BO soliton decay in this case is
similar to that obtained earlier for a KdV soliton
\cite{Grimshaw-2001}. In both these cases the characteristic time
of soliton decay $\tau$ depends only on the initial amplitude and
dissipation parameter $\delta$, whereas in some other cases (e.g.,
in the cases of Reynolds dissipation or Landau damping) it depends
also on the nonlinear, $\alpha$, and dispersion, $\beta$,
coefficients.

\section{Influence of large-scale dispersion on soliton decay}
\label{Sect7}%

In this section we consider the effect of large scale dispersion
on the dynamics of BO solitons. The BO equation augmented by
large scale dispersion was derived by Grimshaw
\cite{Grimshaw-1985} to describe internal waves in a rotating
fluid. It can be presented in a form similar to the Ostrovsky
equation \cite{Grimshaw-1998,Ostrovsky-1978} derived for shallow
water waves
\begin{equation}%
\label{Eq4.1}%
\frac{\partial u}{\partial t} + \alpha u \frac{\partial
u}{\partial x} + \frac{\beta}{\pi} \frac{\partial^2}{\partial
x^2}\,\wp\!\!\int\limits_{-\infty}^{+\infty}\frac{u(\xi, t)}{\xi -
x}d\xi = \gamma \upsilon, \quad u = \frac{\partial
\upsilon}{\partial x}, \quad \upsilon = -\int\limits_x^\infty u(x,
t)\,dx\,.
\end{equation}
Here, $\gamma$ is a parameter characterising the large scale
dispersion.  In the context of internal waves in a rotating ocean
$\gamma = f^2/2c$, where $f$ is the Coriolis parameter (see
\cite{Apel-2007,Grimshaw-1985}).

For the derivation of the energy balance equation we multiply the
perturbed BO Eq.\ (\ref{Eq4.1}) by $u$ and integrate over $x$,
assuming that ahead of the soliton there is no wave
perturbation, i.e.\ $u(x, t) \to 0$ as $x \to \infty$.  We then
obtain
\begin{equation}%
\label{Eq4.2}%
\frac{dE}{dt} = -\gamma \int\limits_{-\infty}^{+\infty}u(x,
t)\left(\int\limits_{-\infty}^{x}u(\xi, t)\,d\xi + C\right)\,dx,
\end{equation}
where $C$ is chosen such that the aforementioned condition at $x = \infty$
is satisfied. Substituting the soliton solution
(\ref{Eq02}) and choosing $C = -\pi/2$, after simple manipulations
we obtain
\begin{equation}%
\label{Eq4.3}%
\frac{dA}{dt} = -\frac{8\pi\gamma\beta}{\alpha},
\end{equation}
which gives
\begin{equation}%
\label{Eq4.4}%
A(t) = A_0\left(1 - \frac{t}{\tau}\right), \quad \tau =
\frac{\alpha A_0}{8\pi\beta\gamma} = \frac{1}{2\pi\gamma\Delta_0},
\end{equation}
where $\Delta_0$ is the initial soliton width.

\begin{figure}
\centering %
%\vspace*{-1.0cm}%
\includegraphics[width=13.5cm]{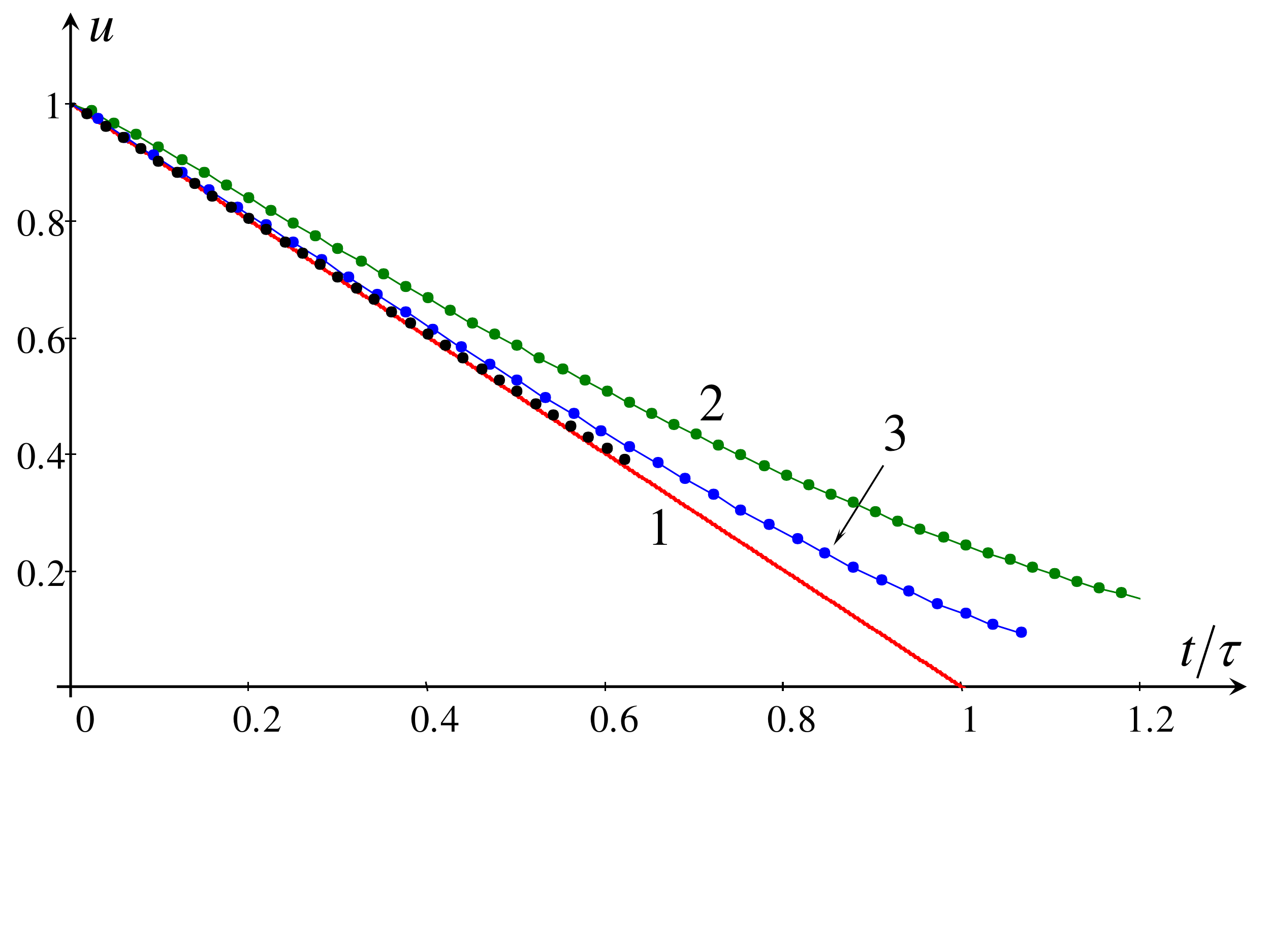}
\vspace*{-2.0cm}%
\caption{(colour online). Soliton amplitude dependence on the
normalised time $t/\tau$ for the radiative decay caused by
large-scale dispersion (line 1). Symbols represent numerical data
obtained with $\alpha = 2$, $\beta = 1$, and $\gamma = 10^{-4}$
(line 2), $\gamma = 10^{-5}$ (line 3), and $\gamma = 10^{-6}$
(dots on line 1).} %
\label{f09}
\end{figure}

The dependence of the amplitude $A(t)$ on time is depicted in
Fig.\ \ref{f09} in which the asymptotic result (\ref{Eq4.4}) is
compared with numerical solutions for various values of the
parameter $\gamma$.  This figure demonstrates that the
applicability of the adiabatic theory is very sensitive to the
value of this parameter, with small values being required for good
agreement with the asymptotic result.  As $\gamma$ decreases, the
time required for significant decay greatly increases as the
half-life of the soliton increases.  Indeed, for $\gamma =
10^{-6}$, the numerical results shown in Figure \ref{f09} are up
to $t=50,000$.

A simple estimate of the relative strength of the large scale
dispersive term ($\sim \gamma \upsilon \sim \gamma A(t)\Delta(t)$)
in comparison with the small scale dispersive term ($\sim \beta
A(t)/\Delta^2(t)$) shows that their ratio is proportional to
\begin{equation}
\frac{\gamma}{\beta}\left(\frac{4\beta}{\alpha A(t)}\right)^3 =
\frac{\gamma}{\beta}\left(\frac{4\beta}{\alpha
A_0}\right)^3\frac{1}{\left(1 - t/\tau\right)^3}.
\end{equation}
This ratio increases with time and leads to the breakdown of the
asymptotic theory. Formally this happens when
\begin{equation}
\frac{t}{\tau} \approx 1 - \frac{4\beta}{\alpha
A_0}\left(\frac{\gamma}{\beta}\right)^{1/3}.
\end{equation}
For $\gamma = 10^{-4}$, $t/\tau \approx 0.91$, whereas
for $\gamma = 10^{-5}$, $t/\tau \approx 0.96$ (the other
parameters are as in Fig.\ \ref{f09}).

\begin{figure}[h!]
\centering %
%\vspace*{-0.5cm}%
\includegraphics[width=12cm]{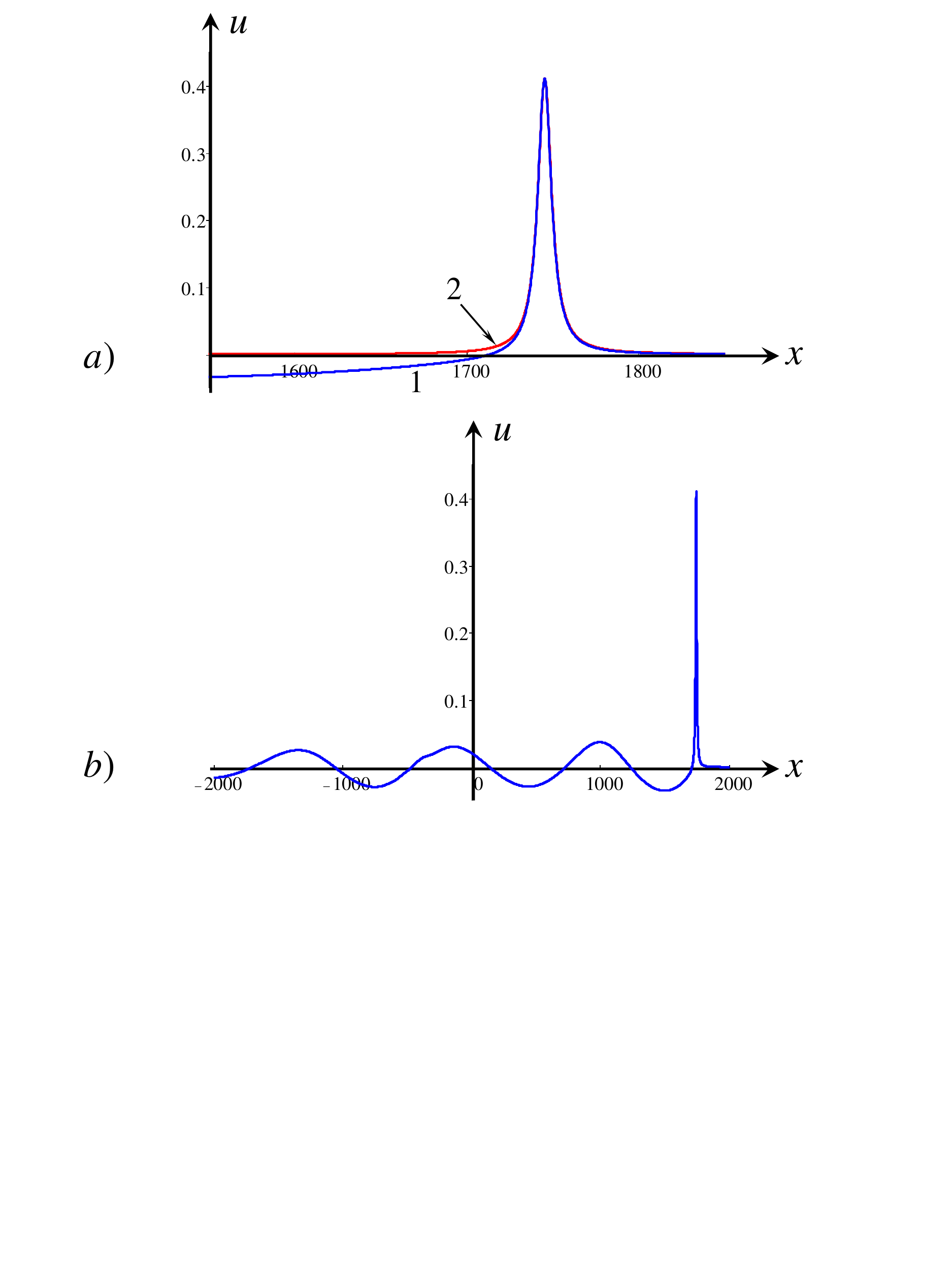}
\vspace*{-6.0cm}%
\caption{(colour online). Soliton profile at $t = 5,000$ ($t/\tau
= 0.63$) for adiabatic decay under the influence of radiative
losses within the framework of Eq.\ (\ref{Eq4.1}) with $\gamma =
10^{-5}$, $\alpha = 2$ and $\beta = 1$. (a) leading solitary wave
(line 1) compared with a BO soliton of the same amplitude (line
2); (b) leading solitary wave shown together with the generated
periodic trailing wave.} \label{f10}
\end{figure}

Note that in the case of the KdV equation the soliton amplitude
also decays to zero in a finite time, but in accordance with the
quadratic law $A(t) = A_0\left(1 - t/\tau\right)^2$
\cite{GrimHeOstr-1998, Grimshaw-1998}. As can be seen from Fig.\
\ref{f09}, the numerical solutions deviate from the theoretical
prediction and the extent of the difference depends on the
parameter $\gamma$. Even for the extremely small value of $\gamma
= 10^{-5}$ the deviation is notable and gradually increases with
time (see line 3).  Figure \ref{f10} illustrates a soliton profile
in the process of adiabatic decay. The soliton is accompanied by
the radiation of a large scale trailing wave.  A similar structure
has been found for a KdV soliton \cite{GrimHeOstr-1998}, which is
expected.

The structure of the trailing wave seen in Fig.\ \ref{f10}(b) can be
calculated with the help of asymptotic theory \cite{GrimHeOstr-1998}.
To this end, let us present a solution of Eq.\ (\ref{Eq4.1}) as an asymptotic
expansion, assuming that $\gamma $ is a small parameter
\begin{equation}
\label{asymp} %
u = u_0 + u_1 + \ldots \,,%
\end{equation} %
where $u_0 $ is a BO soliton (\ref{Eq02}) travelling at speed $V$
in the $x$ direction and $u_1 $ is of order of $\gamma $.
Substituting this expansion into the damped BO Eq.\ (\ref{Eq4.1}), we obtain at
leading order in the small parameter $\gamma$
\be%
\label{exp} %
-V\frac{\partial u_1}{\partial X} + \alpha \frac{\partial
(u_0u_1)}{\partial X} + \frac{\beta}{\pi}
\frac{\partial^2}{\partial
X^2}\,\wp\!\!\int\limits_{-\infty}^{\infty}\frac{u_1(\xi, t)}{\xi
- x}d\xi = \gamma \upsilon_0 - \frac{\partial u_0}{\partial T}\,,
\quad \upsilon_0 = -\int\limits^{\infty }_{X} u_0 \, dx
\,. %
\ee %
Here $\partial u_{0}/\partial T$ is the slow time
evolution of the soliton
\be%
\label{sol} %
u_0(X, T) = \frac{A(T)}{1 + X^2/\Delta^2(T) } \,, \quad X = x -
P(t) \,, \quad \frac{\partial P}{\partial T} = V(T) = \frac{\alpha
A(T)}{4} = \beta\Delta(T)\,. %
\ee
The homogeneous equation for Eq.\ (\ref{exp})
has a solution $u_1 = \partial u_{0}/\partial X$.  The necessary
compatibility condition to remove the secular growing term at this order is then
\be%
\label{comp} %
\int\limits^{\infty }_{-\infty }\, \left(\gamma \upsilon_0 -
\frac{\partial
u_{0}}{\partial T} \right)u_0 \, dX = 0 \,.%
\ee
This yields
\be%
\label{en0} %
\frac{dE_{s}}{dT} = -\gamma \left[\upsilon_{0}^2 \right]_{-\infty } \,, %
\ee %
where $E_s$ is the soliton energy.  Of course, this equation is the same as
Eq.\ (\ref{Eq4.2}).  Therefore, at leading order in the parameter $\gamma$ we obtain Eq.\
(\ref{Eq4.4}) for the soliton amplitude.

Then, bearing in mind that a soliton vanishes at infinity, we
obtain from Eq.\ (\ref{exp}) that as $X \to -\infty $,
\be%
\label{limit1}%
-V\frac{\partial u_{1}}{\partial X} \approx \gamma \upsilon_{0}
\,, \quad \upsilon_{0} \approx -\pi A\Delta = - \frac{4\pi \beta }{\alpha } %
\ee%
in the reference frame moving with the soliton velocity $V$.
It then follows from Eq.\ (\ref{limit1}) that the solution for $u_1$ grows secularly
as $x \to -\infty$
\be%
\label{limit2} %
u_1 \sim \frac{16 \pi \beta \gamma}{\alpha^2 A}X \,. %
\ee
This shelf term can then be matched at $X < 0$ to a trailing sinusoidal wave of
wavenumber $\kappa $,
\be%
\label{sinus} %
u_{shelf} = A_{shelf} \sin{(\kappa X)} \,,
\ee%
where $\kappa^2 = \gamma/V$ and $A_{shelf} \kappa = 16\pi
\beta\gamma/\alpha^2 A$.  Therefore $A_{shelf} =
(8\pi\beta/\alpha^2)\sqrt{\alpha \gamma/A}$.

\begin{figure}[h!]
\centering %
\includegraphics[width=12cm]{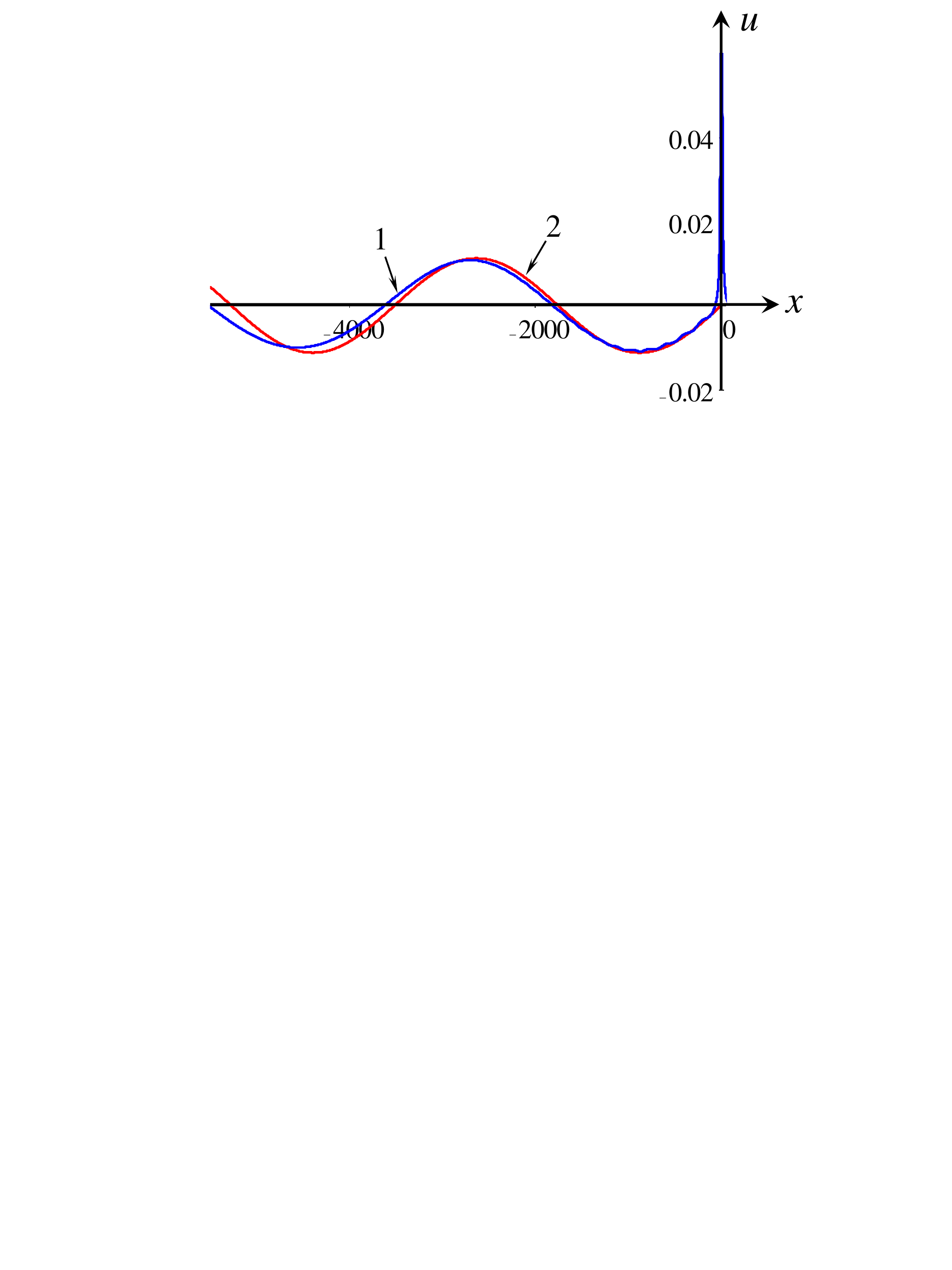}
\vspace*{-11.0cm}%
\caption{(colour online) Trailing wave behind a solitary wave for
large scale dispersion as obtained from numerical calculations
(line 1) and theoretically predicted by Eq.\ (\ref{sinus}) (line
2) at $t = 30,000$ ($t/\tau = 0.377$). The parameters are $\alpha
= 2$, $\beta = 1$ and $\gamma = 10^{-6}$.} %
\label{f11}
\end{figure}

Note that as the main wave decays, this shelf grows, eventually
invalidating the asymptotic expansion (\ref{asymp}) when $\alpha A
\sim \sqrt[3]{(8\pi\beta)^2\gamma}$. Thus the adiabatic theory
breaks down quite quickly. The approximate shelf solution
(\ref{sinus}) is valid in the intermediate stage of soliton decay
when the radiation field is already well formed after a transient
period, but the solitary wave amplitude is still much greater than
the amplitude of the trailing wave, $\alpha A \gg
4\sqrt[3]{\pi^2\beta^2\gamma}$. For the parameters used in Fig.\
\ref{f09}, this means that $A \gg 0.09$, which is in agreement
with the comparisons shown.  Figure \ref{f11} illustrates
comparisons of the wave field behind the leading solitary wave as
obtained in numerical calculations (line 1) and that from Eq.\
(\ref{sinus}) (line 2). The adiabatic theory gives that the
solitary wave completely vanishes in a finite time $\tau$, see
(\ref{Eq4.4}).  However, the long term behaviour is that the
leading wave transforms into a quasi-sinusoidal wavetrain which
can be described by the generalised nonlinear Schr\"{o}dinger
(NLS) equation \cite{Grimshaw-2008, Grimshaw-2012}.  We shall
derive this equation below and give its stationary solution in the
form of an envelope soliton.

\subsection{Extended nonlinear Schr\"{o}dinger equation}
\label{Subsect7.1}%

The nonlinear Schr\"{o}dinger equation is a weakly nonlinear equation describing the
interaction of a wave packet grouped around a central wavenumber.  As such, it arises
in a large number of physical applications, including fluid mechanics and nonlinear optics
\cite{Ablowitz-1981,Karpman-1973, Whitham-1974}.  This equation is
\begin{equation}%
\label{nls}%
i\left(\frac{\partial A}{\partial t}  + c_g \frac{\partial A}{\partial x}\right)
+ \mu \frac{\partial^{2} A}{\partial x^{2}} + \nu |A|^2 A = 0 \,.
\end{equation}
This equation is for the complex envelope $A(x,t)$ of the
weakly nonlinear solution
\begin{equation}%
\label{exp1} %
u = A(x,t)\exp{(ik-i\omega t)} + \hbox{c.c.} + \cdots
\,,
\end{equation}
where $\hbox{c.c.}$ stands for the complex conjugate. Here $\omega
(k) = -\beta k^2 + \gamma/k$ is the linear dispersion relation.
At leading order the envelope $A$ moves with the linear group
velocity $c_g = d\omega/dk = -2\beta k - \gamma /k^2$. The
dispersive term in Eq.\ (\ref{nls}) generically has the
coefficient $\mu = (1/2)(dc_{g}/dk) $, but the coefficient $\nu $
of the cubic nonlinear term is system dependent.

Our concern here is with the situation when $\mu = 0$, selecting a
wavenumber $k = k_m $ for which the group velocity has a local
extremum.  In this case we must replace the standard NLS equation
(\ref{nls}) with the generalised one including the next higher order
(cubic) dispersion term
\begin{equation}%
\label{nls3}%
i(A_t + c_g A_x ) + \mu A_{xx} +  i\mu_1 A_{xxx} + \nu |A|^2 A = 0
\,,
\end{equation}
where $\mu_1 = -(1/6)(d^2c_{g}/dk^2) \ne 0$ is the coefficient of
this third order linear dispersive term. Note that we retain $\mu$,
even although it may be zero, in order to broaden the parameter
space.  Equations of this type arise in nonlinear optics
\cite{Akhmediev-1997, Kivshar-2003, Malomed-2006} where it has
been found advantageous to include in addition the next order
nonlinear terms.  Thus, we extend Eq.\ (\ref{nls3}) (details
are given in the Appendix) to
\begin{equation}%
\label{enls}%
i\left(A_t + c_g A_x \right) + \mu A_{xx} + i\mu_1 A_{xxx} + \nu
|A|^2 A + i\left(\nu_{1} |A|^2 A_x + \nu_{2} A^2 A^{*}_{x}\right)=
0\,.
\end{equation}
Here the $^{*}$ superscript denotes the complex conjugate. Technically the terms
with coefficients $\nu_{1}$ and $\nu_{2} $ are higher order, but
nevertheless they are needed, as will be shown below.

We seek a solitary wave solution of Eq.\ (\ref{enls}) in the form
\begin{equation}%
\label{sol2}%
A = F(X- Vt)\exp{(i\kappa X - i\sigma t)} \,,  \quad X = x - c_g t
\,,
\end{equation}
where we choose the gauge $\kappa$ and the chirp $\sigma$ so that
\begin{equation}%
\label{gauge}%
\nu + 2 \kappa \nu_{2} = \frac{(\nu_{1} + \nu_{2} )\mu}{3\mu_1}
\,, \quad \sigma = 3\kappa V + 8\mu_1\kappa^3  +
\frac{\mu}{\mu_1}\left(4\mu_1\kappa^2 - V\right) - \frac{2\kappa
\mu^2 }{\mu_1} \,.
\end{equation}
Note that when $\mu = 0, \kappa = -\nu/2\nu_{2} $. Then  $F(X)$ is
real valued and satisfies the equation
\begin{equation}%
\label{Feq}%
\mu_1 F_{XX} - \tilde{V} F +\frac{\nu_{1} + \nu_{2} }{3} F^3 = 0
\,, \quad \mbox{where} \quad \tilde{V} = V + 3\mu_1 \kappa^2 -
\mu\kappa \,.
\end{equation}
This equation has a $sech$ solitary wave solution, provided that
$\tilde{V} > 0$ (a suitable choice of $V$ can always achieve this
condition) and $\mu_1(\nu_{1} + \nu_{2} ) > 0 $. Note that the
higher-order nonlinear terms are needed to obtain this solution.
Explicitly
\begin{equation}%
\label{Fsol}%
F = a \, \hbox{sech}\frac{X}{\Delta} \,, \quad \mbox{where} \quad
\tilde{V}  = \frac{\mu_1}{\Delta^2} = \frac{\left(\nu_{1} +
\nu_{2} \right)a^2}{6} \,.
\end{equation}
The development of these envelope solitons from the initial BO
soliton can be studied numerically through very long term
calculation.  However, it is out of the scope of this paper.

\begin{figure}[b!]
\centering %
%\vspace*{2.0cm}%
\includegraphics[width=10cm]{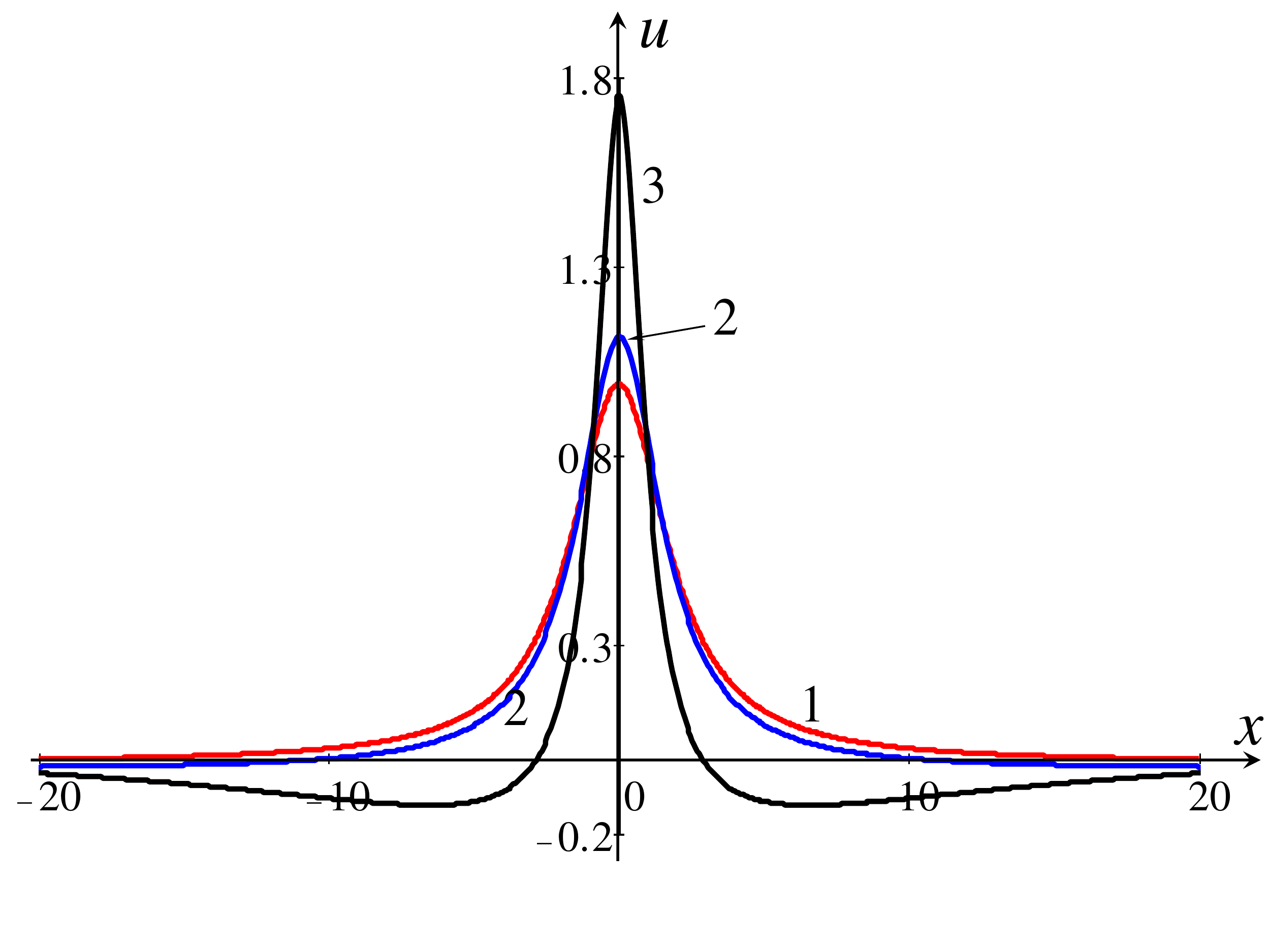}
\vspace*{-1.0cm}%
\caption{(colour online) BO soliton (line 1) compared with
numerical solutions of Eq.\ (\ref{Eq4.1}) for $\gamma = -0.0001$
(line 2) and $\gamma = -0.01$ (line 3). The values of
the other parameters are $\alpha = 2$, $\beta = 1$ and $V = 0.5$.} %
\label{f12}
\end{figure}

To conclude this Section we also consider, for the sake of
completeness, the case of negative $\gamma$, which can be realised
in physical systems with a positive small scale dispersion. In
this case the initial BO soliton does not decay, but gradually
evolves into another soliton of a higher amplitude and zero mean
value.  A similar situation occurs for KdV solitons which
gradually transform into Ostrovsky solitons under the influence of
large scale dispersion \cite{Grimshaw-2016}. In Fig.\ \ref{f12} we
present a stationary soliton obtained numerically using the
Petviashvili method \cite{Petviashvili-1989, Pelinovsky-2004}.
Under the influence of large scale dispersion the soliton becomes
taller and narrower in comparison with the BO soliton.

Thus, within the framework of the asymptotic theory a BO soliton
experiences a terminal decay in accordance with Eq.\ (\ref{Eq4.4})
and completely vanishes in the extinction time $\tau$. However, in
reality it does not completely disappear, but transforms into an
envelope soliton (\ref{Fsol}) after a very long time. This process
is well known for a KdV soliton within the framework of the
Ostrovsky equation \cite{Grimshaw-2008, Grimshaw-2012}.  We do not
consider here the long term evolution of a BO soliton into an
envelope soliton.  This interesting issue will be a matter for a
separate study.

\section{Discussion and conclusion}
\label{Sect8}%

In this paper we have presented a detailed analysis of the
adiabatic decay of a BO soliton under the action of various
dissipation mechanisms. Theoretical solutions derived by means of
asymptotic theory, which presume the smallness of the dissipation,
were compared against numerical solutions. Good agreement between
the theory and numerical solutions was found when the dissipative
coefficients were sufficiently small.  The size of the dissipation
coefficients needed for good agreement depended greatly on the
exact dissipation mechanism.  Estimates of the time scale for the
validity of the asymptotic theory for each type of dissipation
were found.  These were found to be in agreement with numerical
results.  The results of this study are summarised in Table. \\

Table. Summary of cases considered and corresponding decay laws
for a BO soliton

%\smallskip

\begin{table}[h!]
%\noindent
\begin{tabular}{|l|l|l|l|} %
\hline
Type of dissipation & Decay character & Characteristic decay time & Equation \\
\hline %
Rayleigh dissipation & $A(t) = A_0e^{-t/\tau}$ &
$\displaystyle\tau =
\frac{1}{2\delta}$ & Eq.\ (\ref{Eq2.1}) \\
\hline %
Reynolds dissipation & $\displaystyle A(t) = \frac{A_0}{\sqrt{1 +
t/\tau}}$ & $\displaystyle\tau = \frac{8\beta^2}{\delta
\alpha^2A_0^2} = \frac{\Delta_0^2}{2\delta}$ & Eq.\ (\ref{Eq10}) \\
\hline %
Landau damping & $A(t) = \displaystyle\frac{A_0}{1 + t/\tau}$ &
$\displaystyle\tau = \frac{4\beta}{\delta\alpha A_0}
= \frac{\Delta_0}{\delta}$  & Eq.\ (\ref{Eq3.1.5}) \\
\hline %
Decay in a laminar boundary layer & $\displaystyle A(t) =
\frac{A_0}{\left(1 + t/\tau\right)^2}$ & $\displaystyle\tau =
\frac{8}{\delta}\sqrt{\frac{\beta}{\pi\alpha A_0}} =
\frac{4}{\delta}\sqrt{\frac{\Delta_0}{\pi}}$
& Eq.\ (\ref{Eq3.2.5}) \\
\hline%
Chezy friction & $\displaystyle A(t) = \frac{A_0}{1 + t/\tau}$
& $\displaystyle\tau = \frac{2}{3\delta A_0} = \frac{\alpha\Delta_0}{6\beta\delta}$ & Eq.\ (\ref{Eq3.3.2}) \\
\hline %
Radiative decay & $\displaystyle A(t) = A_0\left(1 -
\frac{t}{\tau}\right)$ & $\displaystyle\tau = \frac{\alpha
A_0}{8\pi\beta\gamma} = \frac{1}{2\pi\gamma\Delta_0}$
& Eq.\ (\ref{Eq4.4}) \\
\hline
\end{tabular}
%\caption{Summary of cases considered and corresponding decay laws for a BO soliton.}
\label{t:table}
\end{table}

\bigskip

In particular, it has been found that the solution (\ref{Eq3.3.4})
describing soliton decay due to the Landau damping within the
adiabatic theory is actually the exact solution of the BO equation
with the Landau damping (\ref{Eq3.1.1}).

In the case of the radiative decay of a BO soliton under the
influence of large scale dispersion with a positive loss
coefficient $\gamma$ the theory predicts the terminal decay of the
soliton in finite time. This prediction is in broad agreement with
the numerical data, except for times very close to the extinction
time $\tau$.  Numerical results show that the BO soliton does not
completely vanish, but transforms into an envelope soliton which
can be described by an extended NLS equation. Such an NLS equation
was derived and its soliton solution was obtained in analytic
form.

\bigskip
\bigskip

{\bf Acknowledgements.} The research of R.G. was supported by the
Leverhulme Trust, grant No: EM-2015-037, through the award of a
Leverhulme Emeritus Fellowship. Y.S. acknowledges the funding of
this study from the State task program in the sphere of scientific
activity of the Ministry of Education and Science of the Russian
Federation (Project No. 5.1246.2017/4.6) and grant of the
President of the Russian Federation for state support of leading
scientific schools of the Russian Federation (NSH-2685.2018.5).

\section*{Appendix. Derivation of extended nonlinear Schr\"{o}dinger equation}
\label{Sect9}%

Here we derive the extended NLS equation (\ref{enls}) from the perturbed BO
equation (\ref{Eq4.1}) with large scale dispersive loss.  The envelope solitary
wave solution (\ref{Fsol}) in terms of the coefficients of equation (\ref{Eq4.1})
will then be found. To this end we seek a solution as the asymptotic expansion
\be%
\label{expO} %
u = Ae^{i\theta} + \hbox{c.c.} +
A_2 e^{2i\theta} + \hbox{c.c.}  + A_0 + \ldots \,, %
\ee %
where the phase is $\theta = kx-\omega t$. Here, it is understood that $A(x,t)$
is a slowly varying function of $x$ and $t$.  We expect that
the second harmonic $A_2$ and the zero harmonic (the mean flow term)
$A_0$ are of $O(|A|^2)$, where $|A| \ll 1$, and are likewise slowly varying.

Substituting the expansion (\ref{expO}) into the perturbed BO equation (\ref{Eq4.1}),
we find the dispersion relation
\begin{equation}%
\label{dispersion}%
D\left(\omega,k \right) \equiv \omega k +\beta k^2 |k| -\gamma = 0
\end{equation}
at leading order.  Then the leading order terms in the coefficient of the first
harmonic yields
\be%
\label{firsth} %
D\left(\omega + i\frac{\partial }{\partial t}, \; k -
i\frac{\partial }{\partial x}\right) A  + \left(k
-i\frac{\partial }{\partial x}\right)\, \hbox{NL} = 0 \,, %
\ee%
where the nonlinear term $\hbox{NL}$ is given by
\be%
\label{NL}%
\hbox{NL}  =  - \alpha \left(k-i\frac{\partial}{\partial x}\right)
\left(A_2 A^{*}  + A_0 A  + \ldots \right) \,. %
\ee %
Expanding and noting that $D_{\omega} = k$ gives
\be%
\label{enlsO}%
i\left(A_t + c_g A_x \right)+ \mu A_{xx} + i\mu_1 A_{xxx} +
\hbox{NL} + \ldots = 0  \,, %
\ee %
where $\mu = (1/2)\left(dc_{g}/dk\right)$ and $\mu_1 =
-(1/6)\left(d^2c_{g}/dk^2\right)$, that is
\be%
\label{coeff} %
\mu = \hbox{sign}(k)\left(-\beta + \frac{\gamma }{|k|^ 3}
\right)\, (=0)\,, \quad \mu_1 = \frac{\gamma }{k^4 }
 \left(= \left( \frac{\beta^{4}}{\gamma} \right)^{1/3}\right) \,.%
\ee %
The values in the brackets are those for $|k|= k_m \equiv
\sqrt[3]{\beta/\gamma}$.

It now only remains to find the nonlinear term NL.
First, we note that to leading order
\be%
\label{mean} %
\gamma A_0 - \frac{\partial^2 A_{0}}{\partial x \partial t} +
\cdots = \alpha \frac{\partial^2 |A|^{2}}{\partial x^2} + \ldots%
\ee %
Hence, $A_0 $ is two orders of magnitude smaller than $|A|^2 $,
provided that $\gamma \ne 0$, so that it can be neglected unless $\gamma \ne 0$.
We note that if $\gamma = 0$, then $A_0 = \alpha |A|^2
/c_g$ and it is of the same order as $|A|^2  $.  Also, if
$\gamma \ll 1$, the solution for $A_0 $ is again of the same
order as $|A|^2 $, but is nonlocal.  Henceforth we assume that
$\gamma \ne 0$ and of $O(1)$, since then $k_m $ is also $O(1)$.
Next
\be%
\label{second} %
D\left(2\omega + i\frac{\partial }{\partial t}, 2k -
i\frac{\partial }{\partial x} \right) A_2 = \frac{\alpha}{2}
\left(2k -i \frac{\partial }{\partial x}\right)^2 A^2 \,. %
\ee
At leading order
\be%
\label{second0}%
D_2 A_2 = 2\alpha k^2 A^2 \,, \quad \hbox{where} \quad D_2 =
D\left(2\omega, 2k\right) = 4\beta k^2 |k| + 3\gamma \; (=7\gamma ) \,. %
\ee
As in (\ref{coeff}), the term in brackets is the value at criticality.  As we
need the next order terms as well, we expand to next order to yield
\be%
\label{secpnd1} %
D_2 A_2 = 2 \alpha k^2 A^2 -2i \alpha k\frac{\partial
A^2}{\partial x} - iD_{\omega }(2\omega, 2k)\frac{\partial
A_{2}}{\partial t} + iD_{k}(2\omega, 2k)\frac{\partial
A_{2}}{\partial x} + \ldots  %
\ee %
In the last two terms on the right hand side we may now substitute the leading order
expression for $A_2 $, resulting in
\begin{equation}
\label{second2}%
D_2 A_2 =  2 \alpha k^2 A^2 + iC_2 \frac{\partial
A^2}{\partial x} + \ldots \,,%
\end{equation}
where
\begin{equation}
\label{second3} %
C_2 = -2\alpha k  + 4\alpha k^3 \frac{c_g (k) -
c_g (2k)}{D_2} = - \frac{\alpha k \left(2\beta k^2 |k| + 9\gamma
\right)}{4\beta k^2 |k| + 3\gamma } \; \left( =  -
\frac{11}{7}\alpha k_m\right)\,. %
\end{equation}
Hence, on substituting (\ref{second2}) for $A_{2}$ into Eq.\ (\ref{NL}) for the nonlinear
term we find that
\be%
\label{NL1} %
\hbox{NL} =  - \alpha  k A_2 A^{*} + i\alpha \frac{\partial
\left(A_2 A^{*}\right)}{\partial x} + \ldots  =   \nu |A|^2 A +
i\nu_1 |A|^2 \frac{\partial A}{\partial x} + i\nu_2 A^2
\frac{\partial A^{*}}{\partial x} \,,
\end{equation}
with $\nu  =  -2\alpha^2 k^3/D_2$.

Note that $\hbox{sign}(k)\nu < 0$ for all $k$ and so the NLS
equation (\ref{enlsO}) is focussing for $|k| > k_m$ and defocussing for
$|k| < k_m $, where $k_m$ was defined at the extremum for which $\mu = 0$.
This is in sharp contrast to the reduction of the KdV equation to the NLS equation
for which the NLS equation is always defocussing. The reason is twofold.  First, the presence
of rotation has suppressed the mean flow, and hence changed the
sign of $\nu $ from the KdV case.  Secondly, it has introduced
the critical turning point $|k|=k_m $.  The higher order coefficients are given by
\begin{equation}%
\label{alpha}%
\nu_1 = -\frac{2\alpha k C_2 }{D_2 } + \frac{4\alpha^2 k^2 }{D_2 }
= \frac{2\alpha^2 k^2 \left(10\beta  k^2  |k| + 15 \gamma
\right)}{\left(4\beta k^2 |k| + 3\gamma \right)^2 } \; \left(=
\frac{25\alpha^2 k_{m}^2 }{49 \gamma}\right)
\,,%
\end{equation}
\begin{equation}%
\label{beta}%
\nu_2 = \frac{2 \alpha^2 k^2 }{D_2 } \; \left(= \frac{2 \alpha^2
k_{c}^2 }{7\gamma }\right) \,.
\end{equation}
Thus, both $\nu_{1,2} > 0$ for all $k$, confirming that an envelope
solitary wave given by Eq.\ (\ref{Fsol}) exists for $k=k_m $.

%\newpage

\end{document}